\documentclass[%
reprint,
superscriptaddress,
aps,
pra,
notitlepage,
longbibliography,
]{revtex4-1}

\usepackage[utf8]{inputenc}

\usepackage{graphicx}
\usepackage{dcolumn}
\usepackage{bm}
\usepackage{mhchem}
\usepackage{xcolor}

\usepackage{natbib}
\usepackage[breaklinks,colorlinks=false,linkcolor=blue,urlcolor=blue,citecolor=red]{hyperref}

\usepackage{amsmath, esint, bbold, gensymb, amsfonts, amssymb, bm}

\usepackage{physics}
\usepackage{simpler-wick}
\usepackage{slashed} 
\usepackage[framemethod = tikz]{mdframed}
\usepackage[compat=1.1.0]{tikz-feynman}
\usepackage{tensor}
\usepackage{mhchem}

\newcommand{\hc}{\text{h.c.}}

\renewcommand{\c}{\hat{c}}

\renewcommand{\k}{\textbf{k}}

\renewcommand{\d}{\text{d}}

\DeclareMathOperator{\algsu}{\mathfrak{su}}

\begin{document}

\title{Rise and Fall of Non-Fermi Liquid Fixed Points in Multipolar Kondo Problems}

\author{Daniel J. Schultz}
\affiliation{%
Department of Physics and Centre for Quantum Materials, University of Toronto, Toronto, Ontario, M5S 1A7, Canada
}%

\author{Adarsh S. Patri}
\affiliation{%
Department of Physics and Centre for Quantum Materials, University of Toronto, Toronto, Ontario, M5S 1A7, Canada
}%

\author{Yong Baek Kim}
\affiliation{%
Department of Physics and Centre for Quantum Materials, University of Toronto, Toronto, Ontario, M5S 1A7, Canada
}%

\date{\today}

\begin{abstract}
Recently it was shown that the multipolar Kondo problem, wherein a quantum impurity carrying higher-rank multipolar moments interacts with conduction electrons, leads to novel non-Fermi liquid states. Because of the multipolar character of the local moments, the form of the interaction with conduction electrons is strongly dependent on the orbital-symmetry of the conduction electrons via crystalline symmetry constraints. This suggests that there may exist a variety of different non-Fermi liquid states in generic multipolar Kondo problems depending on the character of conduction electrons. In this work, using renormalization group analysis, we investigate a model where the multipolar local moment is coupled to conduction electrons with two different orbital-symmetry components, namely $p$-wave and $f$-wave symmetries. When each orbital-symmetry component is present alone, non-Fermi liquid states with exactly the same thermodynamic singularities appear. When both orbital-symmetry components are allowed, however, a completely different non-Fermi liquid state arises via the quantum fluctuations in the mixed scattering channels. This remarkable result suggests that the multipolar Kondo problem presents novel opportunities for the discovery of unexpected non-Fermi liquid states.

\end{abstract}

\maketitle

\section{\label{sec:introduction}Introduction}

Classification of non-Fermi liquid states may hold a key for understanding unconventional metallic and superconducting phases in strongly interacting electron/fermion systems \cite{Keimer2015a, Si2016, Chubukov2012, Coleman2017}. This is analogous to the standard paradigm of understanding a plethora of broken-symmetry states as an instability of the Fermi liquid ground state of weakly interacting electron systems. While Fermi liquid theory is based on well-defined quasiparticles, non-Fermi liquids are broadly characterized by the absence of such quasiparticles and the associated singular thermodynamic signatures \cite{Lee2018a}. Hence it is conceivable that there may exist a variety of different non-Fermi liquid states, which may be responsible for unconventional behaviors of metallic, superconducting, and quantum critical regimes of correlated quantum matter including high $T_c$ cuprates, heavy fermions, and various two-dimensional materials with narrow bands \cite{Stewart2001, Cao2020}. It is thus important to understand possibly different origins of non-Fermi liquid ground states.

The conventional Kondo problem, wherein a single magnetic impurity interacts with conduction electron spins, has been a fruitful playground for non-Fermi liquid physics. In particular, non-Fermi liquid ground states arise when a number of channels of conduction electrons greater than or equal to two is coupled to a spin-1/2 impurity \cite{Ludwig1994, Andrei1983, Tsvelick1984, Tsvelick1985}. This valuable lesson, however, has been largely limited to the cases when the local moment only carries a dipole moment, which interacts with conduction electron spins. In many $f$-electron systems, for example, the local moments carry higher-rank multipolar moments \cite{Onimaru2011, Tsujimoto2014, Sato2012, Sakai2012, Riggs2015, Matsubayashi2012b, Rosenberg2019a, Freyer2018, Lee2018b, Patri2019c} and they do not interact solely with the spin, but with more complicated bilinear operators of electrons. Earlier studies in the case of the quadrupolar local moment indeed found a non-Fermi liquid state in both the multipolar lattice setting as well as in the single-multipolar moment (impurity) limit via dilution of the $f$-electron ions \cite{Kuramoto2009, Cox1987, Cox1988, Yamane2018, Yanagisawa2019, Araki2014}.
However, the full generality of the multipolar Kondo problem has not been thoroughly investigated. Recently, some of us have shown that a novel non-Fermi liquid state arises in the multipolar Kondo problem in cubic systems, where the local moment carries quadrupolar and octupolar moments, which interact with conduction electrons with $p$-wave orbital-symmetry ($T_2$ representation of the local $T_d$ symmetry) \cite{Patri2020b,Patri2020c}. This novel non-Fermi liquid state is distinct from the well-known multichannel Kondo non-Fermi liquid states. It was also shown that it is still stable even when additional conduction electrons with $e_g$ orbital-symmetry ($E$ representation of the local $T_d$ symmetry) are introduced. From the point of view towards the classification of non-Fermi liquid states, an important question is how one could control (understand) the emergence (origin) of different kinds of non-Fermi liquid states in the multipolar Kondo problem.

In this work, we consider the multipolar Kondo problem in cubic systems, where the local moment with quadrupolar and octupolar moments interacts with conduction electrons with both $p$-wave ($T_2$ representation of the local $T_d$ symmetry) and $f$-wave ($T_1$ representation of the local $T_d$ symmetry) orbital-symmetries. We use renormalization group (RG) analysis to investigate the presence of non-Fermi liquid fixed points. When only the conduction electrons with $f$-wave orbital-symmetry are present, it is shown that the emergent non-Fermi liquid fixed points are characterized by the same thermodynamic singularities as those of the non-Fermi liquid fixed point found earlier in the $p$-wave model. Hence when the conduction electrons with $p$-wave and $f$-wave orbital-symmetries are separately considered, they lead to the same non-Fermi liquid behaviors. Remarkably, the situation changes dramatically when the conduction electrons with both orbital-symmetries are introduced. This allows the Kondo scattering processes in the mixed orbital-symmetry channels, which leads to the destabilization of the original non-Fermi liquid fixed points and the appearance of a different non-Fermi liquid fixed point. Using RG analysis and a unitary transformation at the fixed point, we show that one of the non-Fermi liquid fixed points is now characterized by a four-channel Kondo non-Fermi liquid behavior. This tantalizing result suggests that the effect of quantum fluctuations, in the presence of multiple symmetry components and their interference, can lead to unexpected non-Fermi liquid fixed points with different thermodynamic behaviors. 
 
The rest of the paper is organized as follows.
In Sec. \ref{sec_micro}, we describe the microscopic constituent degrees of freedom (local multipolar moment and conduction sea) that make up the multipolar Kondo problem.
In Sec. \ref{sec:kondo}, we present the symmetry-permitted multipolar Kondo models with conduction electrons belonging to $T_1$, $T_2$ and $T_1 \otimes T_2$. We also present RG flow equations and the corresponding stable fixed points.
In Sec. \ref{sec_tuned}, we consider the new intermediate fixed point manifold tuned to a special point that provides clarity to the model and the RG flow equations, and present the mapping of the fixed point Hamiltonian to the four-channel Kondo model.
In Sec. \ref{sec_discussions}, we discuss the broader implications of our findings and propose future directions of research.
 
\section{Constituent Degrees of Freedom}
\label{sec_micro}

The combination of spin-orbit coupling and crystalline electric fields places strong constraints on the shape of localized electron wave functions.
This restriction leads to the formation of higher-rank multipolar moments that describe localized anisotropic charge and magnetization densities.
For instance, in the case of a rare-earth Pr$^{3+}$ ion subjected to a surrounding tetrahedral ($T_d$) crystal field, the spin-orbit coupled $J=4$ multiplet of the 4$f^2$ electrons is split to give rise to a low-lying (and well isolated) $\Gamma_{3g}$ non-Kramers doublet \cite{Onimaru2016}.
This $\Gamma_{3g}$ doublet supports both time-reversal even quadrupolar moments $\left\{\hat{\mathcal{O}}_{22} = \frac{\sqrt{3}}{2}(\hat{J}_x^2 - \hat{J}_y^2)\right.$, $\left.\hat{\mathcal{O}}_{20} = \frac{1}{2} (3\hat{J}_z^2 - \hat{\bm{J}}^2)\right\}$ as well as a time-reversal odd octupolar moment $\left\{\hat{\mathcal{T}}_{xyz} = \frac{\sqrt{15}}{6}\overline{\hat{J}_x\hat{J}_y\hat{J}_z}\right\}$; we use the Stevens operators to describe the multipolar moments and the overline indicates a full symmetrization.
These moments can be compactly represented by the pseudospin-1/2 operator $\hat{\bf{S}}$, the components of which
\begin{equation}
\hat{S}^x = \frac{1}{2} \left( \frac{-\hat{\mathcal{O}}_{22}}{4} \right),~~ \hat{S}^y = \frac{1}{2} \left( \frac{-\hat{\mathcal{O}}_{20}}{4} \right), ~~ \hat{S}^z = \frac{1}{2} \left( \frac{\hat{\mathcal{T}}_{xyz}}{3\sqrt{5}}  \right) 
\end{equation}
satisfy a canonically normalized $\algsu(2)$ algebra.
We emphasize that though the multipolar moments are written in terms of pseudospin-1/2 operators, their physical content (and transformations under the symmetry elements) reflects their underlying multipolar nature.

The immersion of multipolar moments in a metallic system permits the local multipolar moment, according to symmetry, to couple to and scatter conduction electrons.
In this work, we focus on scattering conduction electrons belonging to the $T_1$ representation ($\{x(y^2-z^2), y(z^2-x^2), z(x^2-y^2)\}$) of the $T_d$ group in conjunction with the $T_2$ representation ($\{x,y,z\}$) -- which was studied in an earlier work by some of the authors.
In the concrete example of the rare-earth multipolar compounds \ce{Pr(Ti,V)2Al20}, the aforementioned conduction electron states are composed of `molecular orbitals'  formed by a linear combination of the Al atoms' $p$ electrons that surround the Pr$^{3+}$ ion \cite{Nagashima2014, Patri2020b}. The $p$ electrons on the Al atoms are not to be confused with the molecular orbital states in the $T_2$ ($p$-wave symmetry) representation.
We note that these conduction electrons are also equipped with their spin-$1/2$ degree of freedom, which (as will be seen explicitly in the next section) also participates in the Kondo scattering events.

\section{Multipolar Kondo Models}

\label{sec:kondo}

The electrons belonging to the various irreducible representations of $T_d$ participate in scattering events with the impurity. 
In particular, there are intra-irrep and inter-irrep scattering events, where the conduction electrons scatter within basis functions belonging to the same irrep and in different irreps, respectively.
Constrained by the local $T_d$ symmetry about the rare-earth ion and time-reversal symmetry (TRS), we consider the coupling of conduction electron bilinears (possessing orbital and spin degrees of freedom) with the multipolar moments.
We note that in performing the symmetry analysis, it is operationally efficient to express the conduction electron operators in the cubic harmonic basis, and the symmetry operations on said operators are listed in Appendix \ref{app:symmetries}. 
However, since the multipolar moment resides on a site of spin-orbit coupling, it is physically more natural to express the conduction electron operators in terms of a composite spin basis $\ket{j',m_{j'}}\bra{j,m_j}$ as well.
We employ the spin-orbit coupled $j$ basis for all models in this section. 

\subsection{\label{sec:f_model} $T_1$-orbital Kondo model}
The scattering of $T_1$-representation conduction electrons within the $T_1$ irrep corresponds to initial and final orbital angular momenta of $\ell=3$; the subsequent spin-orbit coupled total spin $j$ of these states is either $5/2$ or $7/2$. 
It suffices to state here that only certain linear combinations, Eqs. \eqref{eq:chi1} - \eqref{eq:chi3}, of the composite-spin kets appear in the Kondo Hamiltonian, which we label as $\ket{\chi^{(\pm)}_i}$, where $i=1,2,3$,
\begin{widetext}
\begin{align}
\ket{\chi^{(\pm)}_1} =& \pm\frac{1}{2}\sqrt{\frac{10}{7}}\ket{\frac{5}{2},\frac{\pm 1}{2}} + \frac{1}{4}\sqrt{\frac{30}{7}} \ket{\frac{7}{2},\frac{\pm 1}{2}} - \frac{\sqrt{6}}{4}\ket{\frac{7}{2},\frac{\mp 7}{2}}, \label{eq:chi1} \\
\ket{\chi^{(\pm)}_2} =& \frac{3}{4}\sqrt{\frac{2}{7}} \ket{\frac{7}{2}, \frac{\pm 5}{2}} + \frac{3}{4}\sqrt{\frac{6}{7}} \ket{\frac{7}{2},\frac{\mp 3}{2}} \mp  \frac{5}{6}\sqrt{\frac{3}{7}}\ket{\frac{5}{2},\frac{\pm 5}{2}} \mp  \frac{5}{6}\sqrt{\frac{3}{35}}\ket{\frac{5}{2},\frac{\mp 3}{2}}, \label{eq:chi2} \\
\ket{\chi^{(\pm)}_3} =& \sqrt{\frac{2}{21}} \ket{\frac{5}{2},\frac{\pm 5}{2}} - \sqrt{\frac{10}{21}}\ket{\frac{5}{2},\frac{\mp 3}{2}} \pm  \frac{3}{2\sqrt{7}}\ket{\frac{7}{2},\frac{\pm 5}{2}} \mp \frac{1}{2}\sqrt{\frac{3}{7}} \ket{\frac{7}{2},\frac{\mp 3}{2}}, \label{eq:chi3} 
\end{align}
\end{widetext}
where the $(\pm)$ superscript denotes two time-reversal related pairs of these special linear combinations.

The Kondo interactions in Eqs. \eqref{Hf1}, \eqref{Hf2}, \eqref{Hf3} involve three types of terms with corresponding coupling constants, $F_{Q1}$, $F_{Q2}$, and $F_O$, where their Latin subscripts indicate which multipolar moment is interacting with the conduction electron bilinears i.e. $H^{T_1}_O$ describes interaction with the octupolar moment etc.
%
\begin{widetext}
\begin{align}
H^{T_1}_{Q1} =& F_{Q1}\sum_{s=\pm 1}\left[\hat{S}^x\left\{\ket{\chi^{(s)}_1}\bra{\chi^{(s)}_2} + \hc \right\} + \hat{S}^y\left\{\ket{\chi^{(s)}_1}\bra{\chi^{(s)}_1} - \ket{\chi^{(s)}_2}\bra{\chi^{(s)}_2} \right\} \right] ,\label{Hf1} \\
H^{T_1}_{Q2} =& F_{Q2} \sum_{s=\pm1}\left[\hat{S}^x\left\{s\ket{\chi^{(s)}_1} \bra{\chi^{(s)}_3} \right\} + \hat{S}^y\left\{s\ket{\chi^{(s)}_2} \bra{\chi^{(s)}_3}  \right\} + \hc \right], \label{Hf2} \\
H^{T_1}_O =& F_O \sum_{s=\pm1}\hat{S}^z\left[i\ket{\chi^{(s)}_1} \bra{\chi^{(s)}_2} + \hc\right]. \label{Hf3}
\end{align}
\end{widetext}
We notice the two distinct sectors exhibited in this Hamiltonian: an electron in a $\ket{\chi^{(\pm)}_i}$ state may only transition into another $\ket{\chi^{(\pm)}_{i'}}$ state. The first quantized notation $\ket{\alpha}\bra{\beta}$ is used in place of second quantization $\c^\dagger_\alpha \c_\beta$ for the sole reason of making the would-be subscripts more readable.

\subsection{\label{sec:p_model} $T_2$-orbital Kondo Model}

Scattering of conduction electrons within the $T_2$ representation via the impurity leads to an interaction term which is remarkably identical in form to that of the $T_1$ electrons. 
Indeed, the $T_2$ model was the focus of an earlier work by some of the authors \cite{Patri2020b}, wherein the scattering involving $j=3/2$ and $j=1/2$ electrons gave rise to a novel non-Fermi liquid fixed point.
We present in Table \ref{tb:T1_T2_correspondence} the correspondence of the basis states of the $T_1$ and $T_2$ irreps that allows one to notice the isomorphic form of their corresponding Kondo models.
We present the $T_2$ interaction Hamiltonian (along with its three coupling constants $P_{Q1}$, $P_{Q2}$, and $P_O$) in Appendix \ref{app:T2_model}. 

\begin{table}[h]
\renewcommand{\arraystretch}{1.5}
\begin{tabular}{c|cccccc}
$T_1$ & $\ket{\chi^{(+)}_1}$ & $\ket{\chi^{(+)}_2}$ & $\ket{\chi^{(+)}_3}$ & $\ket{\chi^{(-)}_1}$ & $\ket{\chi^{(-)}_2}$ & $\ket{\chi^{(-)}_3}$ \\ 
 & $\updownarrow$ & $\updownarrow$ & $\updownarrow$ & $\updownarrow$ & $\updownarrow$ & $\updownarrow$ \\
$T_2$ & $\ket{\frac{3}{2},\frac{3}{2}}$ & $\ket{\frac{3}{2},\frac{-1}{2}}$ & $\ket{\frac{1}{2},\frac{-1}{2}}$ & $\ket{\frac{3}{2},\frac{-3}{2}}$ & $\ket{\frac{3}{2},\frac{1}{2}}$ & $\ket{\frac{1}{2},\frac{1}{2}}$ 
\end{tabular}
\caption{Correspondence between $T_1$ and $T_2$ basis states in spin-orbit composite basis for the purposes of constructing the $T_2$ interaction Hamiltonian.} \label{tb:T1_T2_correspondence}
\end{table}

\subsection{$T_1\otimes T_2$ Kondo Model}
Due to the multitude of available orbitals, we now consider electrons transitioning between the $T_1$ and $T_2$ molecular orbitals via interaction with the multipolar impurity. The symmetry constraints introduce 5 further coupling constants which we call $X_{Q1}$, $X_{Q2}$, $X_{Q3}$, $X_{O1}$, and $X_{O2}$. This brings us to a grand total of 11 couplings. 
Each individual operator here brings an electron in a $j=1/2$ or $3/2$ state to one which is in a (superposition of) $j=5/2$ or $7/2$ state(s), or vice versa. 
This explicitly indicates that electrons are switching between states in the $T_1$ or $T_2$ representations. 
In the spin-orbit coupled basis, we observe that electrons in any of the 6 states $\left\{\ket{\chi^{(+)}_i}, \ket{\frac{3}{2},\frac{-3}{2}}, \ket{\frac{3}{2},\frac{1}{2}}, \ket{\frac{1}{2},\frac{1}{2}} \right\}$ never transition to any of the other 6 states $\left\{\ket{\chi^{(-)}_i}, \ket{\frac{3}{2},\frac{3}{2}}, \ket{\frac{3}{2},\frac{-1}{2}}, \ket{\frac{1}{2},\frac{-1}{2}} \right\}$, where we recall that $i=1,2,3$.
This segregation of the scattering conduction electrons into two sectors becomes an important ingredient in understanding the nature of this model's fixed points. The explicit forms of the $T_1\otimes T_2$ Kondo interactions are enumerated in Eqs. \eqref{eq:Hfp1} - \eqref{eq:Hfp5}. 

\begin{widetext}
\begin{align}
H^{T_1\otimes T_2}_{Q1} =& X_{Q1}\sum_{s=\pm 1}\left[ \hat{S}^x\left\{\ket{\frac{3}{2},\frac{s}{2}}\bra{\chi^{(s)}_2} - \ket{\frac{3}{2},\frac{3s}{2}}\bra{\chi^{(-s)}_1}\right\} + \hat{S}^y\left\{\ket{\frac{3}{2},\frac{s}{2}}\bra{\chi^{(s)}_1} + \ket{\frac{3}{2},\frac{3s}{2}}\bra{\chi^{(-s)}_2}\right\} + \hc\right] \label{eq:Hfp1}\\
H^{T_1\otimes T_2}_{Q2} =& X_{Q2} \sum_{s=\pm1}\left[ \hat{S}^x\left\{- s\ket{\frac{1}{2},\frac{s}{2}}\bra{\chi^{(s)}_2} \right\} + \hat{S}^y\left\{ s\ket{\frac{1}{2},\frac{s}{2}}\bra{\chi^{(s)}_1} \right\} + \hc \right] \label{eq:Hfp2} \\
H^{T_1\otimes T_2}_{Q3} =& X_{Q3} \sum_{s=\pm1}\left[ \hat{S}^x\left\{s\ket{\frac{3}{2},\frac{s}{2}}\bra{\chi^{(s)}_3} \right\} + \hat{S}^y\left\{s\ket{\frac{3}{2},\frac{3s}{2}}\bra{\chi^{(s)}_3} \right\} + \hc \right] \label{eq:Hfp3} \\
H^{T_1\otimes T_2}_{O1} =& X_{O1} \sum_{s=\pm1}\hat{S}^z \left[- i\ket{\frac{1}{2},\frac{s}{2}}\bra{\chi^{(s)}_3} + \hc\right] \label{eq:Hfp4} \\
H^{T_1 \otimes T_2}_{O2} =& X_{O2} \sum_{s=\pm1} \hat{S}^z\left[i\ket{\frac{3}{2},\frac{3s}{2}}\bra{\chi^{(-s)}_1}  + i \ket{\frac{3}{2},\frac{s}{2}}\bra{\chi^{(s)}_2} + \hc \right] \label{eq:Hfp5}
\end{align}
\end{widetext}

\section{Renormalization Group (RG) Analysis} 

In Wilsonian RG, the coupling constants explicitly depend on the UV cutoff $D$, which physically corresponds to the conduction electron bandwidth in the multipolar Kondo problem \cite{Wilson1975a}.
In this work, we employ perturbative renormalization group theory whereby the perturbatively computed low-energy scattering rate is taken to be independent of the high-energy cutoff.
This leads to the coupling constants explicitly depending on $D$, and `flowing' as $D$ is lowered.
We present the Feynman diagrams responsible for the RG flow equations in Appendix \ref{app_feynman}.
Of interest in this work are the stable fixed points of the RG equations, which correspond to different low-energy theories.
Indeed, the slope ($\Delta$) of the flow equations about a fixed point is related to the scaling dimension $(1+\Delta)$ of the leading irrelevant operator of that low-energy theory, and as such determines the behavior of physical observables such as resistivity and specific heat capacity.
For clarity, we note that the conduction electron densities of states for the different irreps are in principle different; however, these densities of states are implicitly absorbed into the below couplings to yield dimensionless coupling constants. 
%
%

\subsection{\label{sec:f_stability} $T_{1,2}$ Representation model}
Perturbatively expanding the interaction vertices Eqs. \eqref{Hf1} - \eqref{Hf3} to third order in coupling constant strength results in the following flow equations for the coupling constants:
\begin{align}
\frac{\d F_{Q1}}{\d\log D} =& - 2 F_{O} F_{Q1} + 2 F_{Q1} \left(F_{O}^{2} + F_{Q1}^{2} + F_{Q2}^{2}\right), \label{eq:beta_T1FQ1} \\
\frac{\d F_{Q2}}{\d\log D} =& F_{O} F_{Q2} + 2 F_{Q2} \left(F_{O}^{2} + F_{Q1}^{2} + F_{Q2}^{2}\right),  \label{eq:beta_T1FQ2} \\
\frac{\d F_O}{\d\log D} =& - 2 F_{Q1}^{2} + F_{Q2}^{2} + 4 F_{O} \left(F_{Q1}^{2} + F_{Q2}^{2}\right). \label{eq:beta_T1FO}
\end{align}
There are three kinds of fixed points for the flow equations \eqref{eq:beta_T1FQ1} - \eqref{eq:beta_T1FO}. The first is the trivial Gaussian fixed point, where $G = (F_{Q1}, F_{Q2}, F_O) = (0,0,0)$, which is unstable and is therefore not of our interest. The second is $M = (F_{Q1}, F_{Q2}, F_O) = (\pm\frac{1}{2}, 0, \frac{1}{2})$, which corresponds to the well-known 2-channel Kondo model \cite{Patri2020c}. The last, and potentially novel, fixed point is $N = (F_{Q1}, F_{Q2}, F_O) = (0, \pm\frac{1}{4}, -\frac{1}{4})$. We note that both $M$ and $N$ are nontrivial and stable. 

Since the $T_2$ Kondo Hamiltonians are related to the $T_1$ ones via Table \ref{tb:T1_T2_correspondence}, the $\beta$-functions are identical to Eqs. \eqref{eq:beta_T1FQ1} - \eqref{eq:beta_T1FO} with the replacement of $F_{Q1}$, $F_{Q2}$, and $F_O$ by $P_{Q1}$, $P_{Q2}$, and $P_O$, respectively. 
The corresponding (identical) fixed points $M$ and $N$ have been explored in detail in previous work \cite{Patri2020b, Patri2020c}, where it was shown that $M$ is a 2-channel Kondo problem, and $N$ is a novel point. 
These conclusions hold for the $T_1$ model as well.

\subsection{\label{sec:fp_stability} $T_1\otimes T_2$ Mixing Models}

With the incorporation of inter-irrep scattering, the corresponding flow equations naturally become more complicated.
The fixed point solution sets now vary in dimension, and there exist numerous unstable manifolds of fixed points, but importantly two stable manifolds. 
Each of the two stable manifolds is parametrized by 1 parameter, with the precise forms of the solutions given in Appendix \ref{app:fixed_points}. 
From the scaling dimension of the leading irrelevant operator of each of the stable manifolds, we identify one solution-manifold describing two-channel Kondo behaviors, just like the point $M$ in the individual $T_1$ and $T_2$ models. 
However, the other solution is not a direct extension of the $N$ solution from the individual $T_1$ and $T_2$ cases, but rather yields completely new behavior.
Indeed the original $N$ fixed point becomes unstable with these ``mixing'' terms, which suggests that the $T_2$ orbitals act as a relevant perturbation to the $T_1$ orbitals, to yield a new fixed manifold $L$. 

\section{Nature of emergent intermediate fixed point-manifold $L$}
\label{sec_tuned}

The above perturbative RG analysis discovered a potentially-novel stable fixed point manifold ($L$) parameterized by a single variable. 
Since the scaling behavior of the leading irrelevant operator is identical at any point on the manifold, we focus on a specially tuned point that provides clarity to the model. 
This point, $L^*$, is such that all of the coupling constants, except for $P_{Q2}, P_O, X_{Q3}$, and $X_{O1}$, are conveniently set to zero.
At this point we fix the ratios between the surviving coupling constants to be those at the fixed point: $P_{Q2} = P_O = -X_{Q3} = X_{O1}  = -g/4 $, where when $g \rightarrow 1$ we arrive at the perturbative fixed point $L^*$.
Examining this point provides remarkable insight in the nature of the RG flow and the intermediate fixed point Hamiltonian.

\subsection{Renormalization group flow about intermediate-tuned fixed point}

\begin{figure}[ht]%
\includegraphics{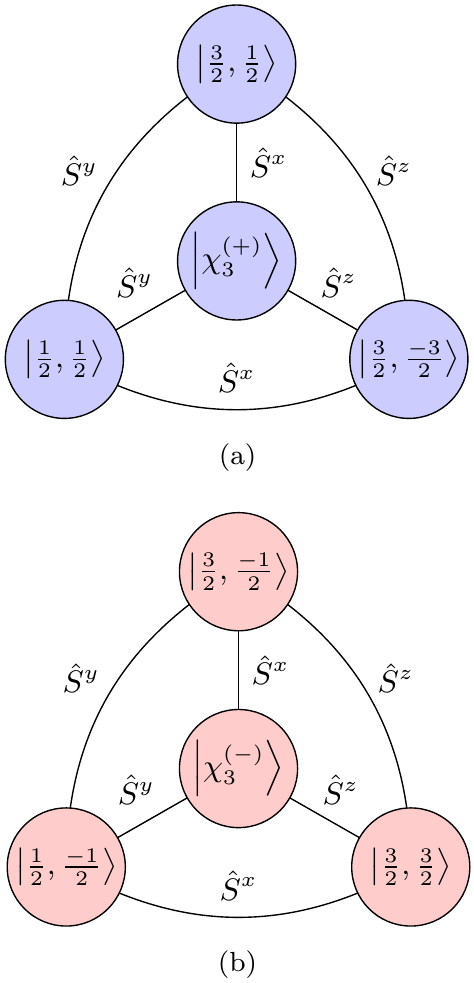}
\caption{The $L$ Hamiltonian tuned to the simplified fixed point $H_{L^*}$ contains 4-states in the $s=\pm$ channels. This figure visualizes the  states and degrees of freedom accessible in $H_{L^*}$. (a) and (b) correspond to the $s=+$ and $s=-$ channels, respectively.}%
\label{fig:ts}%
\end{figure}

With the above parameterization, the fixed point Kondo Hamiltonian arrives at the much simplified form as presented in Eq. \eqref{eq:L_point_model} and schematically in Fig. \ref{fig:ts},
\begin{widetext}
\begin{align}
H_{L^*} =& -\frac{g}{4}\sum_{s=\pm 1} \left[\hat{S}^x\left\{s\ket{\frac{3}{2},\frac{3s}{2}}\bra{\frac{1}{2},\frac{-s}{2}} - s\ket{\frac{3}{2},\frac{s}{2}}\bra{\chi^{(s)}_3} \right\} + \hat{S}^y\left\{s\ket{\frac{3}{2},\frac{-s}{2}}\bra{\frac{1}{2},\frac{-s}{2}} - s\ket{\frac{3}{2},\frac{3s}{2}}\bra{\chi^{(s)}_3} \right\}+ \hc \right] \nonumber \\
&  -\frac{g}{4} \sum_{s=\pm1} \hat{S}^z\left\{i\ket{\frac{3}{2},\frac{-3s}{2}}\bra{\frac{3}{2},\frac{s}{2}} - i\ket{\frac{1}{2},\frac{s}{2}}\bra{\chi^{(s)}_3} + \hc \right\} ,\label{eq:L_point_model}
\end{align}
\end{widetext}
with the corresponding $\beta$-function,
\begin{equation}
\frac{\d g}{\d\log D} = -\frac{g^2}{2} + \frac{g^3}{2}. \label{eq:novel_flow} 
\end{equation}
The scaling dimension ($1+\Delta$) of the leading irrelevant operator can now be easily extracted from the slope of $\beta(g)$ at the fixed point $g^* = 1$ i.e. $\Delta = 1/2$.

\subsection{Mapping of intermediate-fixed point Hamiltonian to four-channel Kondo model}

The form of the Kondo Hamiltonian in Eq. \eqref{eq:L_point_model} possesses an elegant structure: the multipolar impurity couples to two `copies' (or channels) of a four-dimensional manifold of states.
We denote the four-dimensional states of the two channels by $(s=+)$ : $\left\{ \ket{\frac{3}{2}, \frac{-3}{2}},  \ket{\frac{3}{2}, \frac{1}{2}}, \ket{\frac{1}{2}, \frac{1}{2}}, \ket{\chi_3^{(+)}}     \right\}$, and $(s=-)$ : $\left\{ - \ket{\frac{3}{2}, \frac{3}{2}}, - \ket{\frac{3}{2}, \frac{-1}{2}}, \ket{\frac{1}{2}, \frac{-1}{2}}, \ket{\chi_3^{(-)}}     \right\}$, where the minus sign indicates a unitary transformation in the conduction basis.
The Kondo Hamiltonian can then be rewritten using the SU(4) generalized Gell-Mann matrices,
\begin{align}
H_{L^*} =  \frac{g}{2} \sum_{s= \pm} \sum_{\alpha, \beta =1}^{4} &\Big[ \hat{S}^x \left(T^4 - T^{11} \right)_{\alpha \beta} + \hat{S}^y \left(T^6 + T^9 \right)_{\alpha \beta} \Big. \nonumber  \\
& \Big. + \hat{S}^z \left(T^2 + T^{14} \right)_{\alpha \beta} \Big] c_{s, \alpha}^{\dag} c_{s, \beta},
\label{eq_kondo_fp_1}
\end{align}
where $\c_{s, \alpha}^{\dag}$ is a conduction creation operator of spin-orbital state $\alpha$ in the corresponding channel $s$;
$\alpha, \beta$ sum over the two aforementioned four-state sectors, and we use the standard notation for the SU(4) generators, $T$.
Intriguingly, these combinations of SU(4) generators satisfy a canonical $\algsu(2)$ algebra i.e. \newline
$\Big[ T^4 - T^{11},  T^6 + T^9\Big] =  i  \left( T^2 + T^{14} \right) $ etc.
Performing a unitary transformation that diagonalizes $\left(T^2 + T^{14} \right)$ in each channel $s$ rewrites the above Eq. \ref{eq_kondo_fp_1} to,
\begin{align}
H_{L^*} =  \frac{g}{2} \sum_{s= \pm} \sum_{\alpha, \beta =1}^{4} &\Bigg[ \hat{S}^x \left(T^6 + T^{9} \right)_{\alpha \beta} + \hat{S}^y \left(T^7 + T^{10} \right)_{\alpha \beta} \Bigg. \nonumber  \\
& \Bigg. + \hat{S}^z \left( \frac{2}{\sqrt{3}} T^8 + \sqrt{\frac{2}{3}}T^{15} \right)_{\alpha \beta} \Bigg]  \psi_{s, \alpha}^{\dag} \psi_{s, \beta},
\label{eq_kondo_fp_2}
\end{align}
where $\vec{\psi} = U^{\dag} \vec{c}$ is the conduction electron operator in the diagonalized basis, and a canonical transformation has been performed on the multipolar impurity pseudospin operator i.e. $\hat{S}^{x,y} \rightarrow - \hat{S}^{x,y}$.
We list the diagonalized basis in Appendix \ref{app:diagonal_basis}.
Expanding Eq. \eqref{eq_kondo_fp_2} into its components yields
\begin{align}
H_{L^*} =  \frac{g}{2} \sum_{s= \pm} \Big[ & \hat{S}^x \left(  \hat{\psi}_{s, 1}^{\dag} \hat{\psi}_{s, 4} + \hat{\psi}_{s, 4}^{\dag} \hat{\psi}_{s, 1} + \hat{\psi}_{s, 2}^{\dag} \hat{\psi}_{s, 3} + \hat{\psi}_{s, 3}^{\dag} \hat{\psi}_{s, 2}  \right) \Big. \nonumber \\
\Big.  + & \hat{S}^y \left( -i  \hat{\psi}_{s, 1}^{\dag} \hat{\psi}_{s, 4} + i \hat{\psi}_{s, 4}^{\dag} \hat{\psi}_{s, 1} - i  \hat{\psi}_{s, 2}^{\dag} \hat{\psi}_{s, 3} + i\hat{\psi}_{s, 3}^{\dag} \hat{\psi}_{s, 2}  \right)  \Big. \nonumber \\
\Big.  + &  \hat{S}^z \left(  \hat{\psi}_{s, 1}^{\dag} \hat{\psi}_{s, 1} - \hat{\psi}_{s, 4}^{\dag} \hat{\psi}_{s, 4} + \hat{\psi}_{s, 2}^{\dag} \hat{\psi}_{s, 2} - \hat{\psi}_{s, 3}^{\dag} \hat{\psi}_{s, 3}  \right) \Big] \\
= \frac{g}{2} \sum_{k= 1}^{4} & \sum_{\alpha, \beta =1}^{2}  \Big[ \hat{S}^x \frac{\tau^x _{\alpha \beta}}{2}+ \hat{S}^y   \frac{\tau^y _{\alpha \beta}}{2}  + \hat{S}^z   \frac{\tau^z _{\alpha \beta}}{2} \Big]  \hat{\psi}_{k, \alpha}^{\dag} \hat{\psi}_{k, \beta}.
\end{align}
From the first equality, it is apparent that for given channel $s = \pm$, there are two additional channels, which are spanned by states $\{1,4\}$ and $\{2,3\}$, respectively; $\alpha, \beta$ denote the two spin-orbital entangled states in this additional channel, and are expressed by the pseudo-spin operator $\vec{\tau}$ in the second equality.
Remarkably, by examining the second equality, the Kondo model is now identical to that of a four channel-Kondo model.
Drawing upon the conformal field theory (CFT) solution for the four-channel Kondo model \cite{Affleck1993a, Affleck1991}, the exact scaling dimension of the leading irrelevant operator is related to $\Delta_{\text{CFT}} = \frac{2}{k+2} = 1/3$, as $k=4$.
Indeed, the obtained perturbative scaling dimension is consistent with taking the large $k$ limit and setting $k \rightarrow 4$ in $\Delta_{\text{CFT}}$. 
We summarize all fixed point s in Table \ref{tb:scalings}.
We note that though the Kondo model is identical to that of a four-channel Kondo model, the physical content of the conduction electron ``spin'' and ``channel'' are entangled combinations of conduction spin and orbital degrees of freedom, unlike the ordinary four-channel Kondo model where it is purely the spin of the conduction electron participating in the quantum scattering events.

\begin{table}[t]
\begin{tabular}{c|c|c|c}
Model & Fixed Manifold & $\Delta$ (perturbative) & $\Delta$ (exact) \\ \hline
$T_1$-only & 2-Channel $(M)$ & 1 & 1/2\\
& Novel $(N)$ & 1/4 & 1/5 \\ \hline
$T_2$-only & 2-Channel $(M)$ & 1 & 1/2 \\
& Novel $(N)$ & 1/4 & 1/5 \\ \hline
$T_1\otimes T_2$ & 2-Channel $(M)$ & 1 & 1/2 \\
& 4-Channel $(L)$ & 1/2 & 1/3 \\
\end{tabular}
\caption{{Scaling dimension $1+\Delta$ of leading irrelevant operator of different fixed manifolds. The resistivity $\rho$ and specific heat $C$ at low temperatures can be calculated by using $\rho\sim T^\Delta$ and $C\sim T^{2\Delta}$. The $\Delta$ in the third column is from 2-loop order in perturbation theory, and whereas the fourth column scaling is from CFT \cite{Affleck1993a, Affleck1991, Patri2020c}.}} \label{tb:scalings}
\end{table}

\section{Discussions}

\label{sec_discussions}

In this work, we investigated possible non-Fermi liquid states in multipolar Kondo models, where the conduction electrons have two different orbital-symmetry components: $p$-wave ($T_2$ representation) and $f$-wave($T_1$ representation). The model is chosen such that when only one of the orbital-symmetry components is present, the nature of the emergent non-Fermi liquid state is exactly the same. For example, these models separately support the non-Fermi liquid fixed point $N$ described in the main text. However, when both components are present, the mixing between two orbital-symmetry channels via the Kondo scattering processes is allowed. It is shown that, due to the different orbital-symmetries of two components, quantum fluctuations in the mixing channels promotes a completely different non-Fermi liquid fixed manifold, denoted $L$ in the main text, in the renormalization group flow. We identified this fixed manifold $L$ as the four-channel Kondo non-Fermi liquid state.

In the case of multipolar Kondo problems, the type of multipolar moment and symmetries of available conduction electrons depend only on the local symmetry. In many crystal symmetries, such as octahedral or tetragonal, there are no irreducible representations with dimension greater than 3. We have shown here that nontrivial behavior can arise from 3-dimensional irreps, and further nontrivialities can arise by coupling two 3-dimensional irreps together. In apparent contrast, for a single 2-dimensional irrep, e.g. $E$-symmetry orbitals, the fixed point will often produce 2-channel Kondo behavior. This may appear to suggest that no interesting states are present for lower-dimensional irreps. What we have shown is that this is not necessarily the case; it may be that crystals which have numerous 2 dimensional irreducible representations can still support a variety of non-Fermi liquid fixed points via the combination of different 2-dimensional irreps for conduction electrons.

Future directions for the work would be to investigate non-Fermi liquid fixed points in other kinds of multipolar quantum impurity systems. The nature of the impurity, for example the number of electrons on the impurity or the moments supported by the surrounding crystal field, as well as the symmetries of the conduction electron orbitals are all variables for further study. Another direction would be a pursuit of the multipolar Kondo lattice problem. Although a number of experiments have established the existence of non-Fermi liquid behavior in the lattice setting \cite{Sakai2011a}, developing a thorough understanding from a theoretical view is still an active pursuit of the community \cite{Wugalter2020,Inui2020,Lai2018,Hu2020}. Examining the nature of our model in a lattice setting would be an intriguing direction for future work. Our current work places the first piece of the puzzle in the classification of non-Fermi liquid fixed points in multipolar quantum impurity systems. This classification may provide a unified framework for an interesting subset of the strange metal states observed in many modern theoretical and experimental work, leading us towards a general understanding of non-Fermi liquid behavior.


\section*{Acknowledgements}
This work was supported by NSERC of Canada and the Center for Quantum Materials at the University of Toronto. Y.B.K. is supported by the Killam Research Fellowship of the Canada Council for the Arts. 

\appendix

\begin{appendix}

\section{Multipolar Impurity}
The impurity consists of two electrons in the $4f$ configuration. By Hund's rules, the two electrons are $(\ell_1=3,m_1=3,s_1=\downarrow)$ and $(\ell_2 = 3, m_2 = 2, s_2 = \downarrow)$, which leads to a $J=4$ composite spin system as the ground state. In the presence of the $T_d$ crystal field, this $J=4$ representation decomposes into the following irreducible representations $\Gamma_{J=4} = A_{1g} \oplus E_g \oplus T_{1g} \oplus T_{2g}$, where we have used the octahedral group notation for these irreps. The two basis functions for $E_g$ which have $J =4$ are $\{x^4 - y^4, 2z^4 - x^4 - y^4\}$. Expressing these in terms of spherical harmonics, we find that the two degenerate ground states are 

\begin{align}
&\ket{x^4 - y^4} = \ket{\Gamma^{(1)}_3} = \frac{1}{2}\sqrt{\frac{7}{6}}\ket{4} - \frac{1}{2}\sqrt{\frac{5}{3}}\ket{0} + \frac{1}{2}\sqrt{\frac{7}{6}}\ket{-4}, \\
&\ket{2z^4 - x^4 - y^4} = \ket{\Gamma^{(2)}_3} = \frac{1}{\sqrt{2}}\ket{2} + \frac{1}{\sqrt{2}} \ket{-2},
\end{align}

where $\ket{m} = Y^m_4$ is the $\ell=4$ spherical harmonic with $z$-component of angular momentum $m$, and $ \ket{\Gamma^{(1,2)}_3}$ are mutually orthogonal. We therefore define two new states

\begin{align}
\ket{\uparrow} =& \frac{1}{\sqrt{2}}\left(\ket{\Gamma^{(1)}_3} + i\ket{\Gamma^{(2)}_3}\right), \\
\ket{\downarrow} =& \frac{1}{\sqrt{2}}\left(i\ket{\Gamma^{(1)}_3} + \ket{\Gamma_3^{(2)}}\right),
\end{align}

\noindent which, by a quick calculation, are seen to be orthogonal. To determine the supported multipolar moments by these electronic wave functions, we sandwich all the possible Stevens operators between $\ket{\uparrow}$ and $\ket{\downarrow}$.
We determine non-vanishing matrix elements for three Stevens operators in this $2\times 2$ non-Kramers doublet space,

\begin{align}
\hat{\mathcal{O}}_{20} =& \frac{1}{2}(3\hat{J}_z^2 - \hat{\bm{J}}^2) \\
\hat{\mathcal{O}}_{22} =& \frac{\sqrt{3}}{2}(\hat{J}_+^2 + \hat{J}_-^2) \\
\hat{\mathcal{T}}_{xyz} =& \frac{\sqrt{15}}{6}\overline{\hat{J}_x\hat{J}_y\hat{J}_z}.
\end{align}

We clarify that the overline includes an implicit division by $3! =6$, to account for the number of permutations. Additionally, we clarify that the $\hat{J}_x, \hat{J}_y, \hat{J}_z$ operators are $J=4$ spin operators. The resulting $2\times 2$ matrix expressions are

\begin{equation}
\hat{\mathcal{O}}_{22} = -4\sigma^x,\qquad \hat{\mathcal{O}}_{20} = -4\sigma^y,\qquad \hat{\mathcal{T}}_{xyz} = 3\sqrt{5}\sigma^z,
\end{equation}

\noindent where $\sigma^x,\sigma^y,\sigma^z$ are the ordinary Pauli matrices, satisfying $[\frac{\sigma^i}{2},\frac{\sigma^j}{2}] = i\epsilon_{ijk}\frac{\sigma^k}{2}$. To make these Stevens operators, projected into the space of the 2 degenerate ground states (i.e. into pseudo-spin operators) we simply divide by the appropriate coefficients to find that 

\begin{equation}
\hat{S}^x = \frac{1}{2}\left(-\frac{\hat{O}_{22}}{4}\right), \hat{S}^y = \frac{1}{2}\left(-\frac{\hat{O}_{20}}{4}\right),\hat{S}^z = \frac{1}{2}\left(\frac{\hat{\mathcal{T}}_{xyz}}{3\sqrt{5}}\right).
\label{eq_multipole_op}
\end{equation}
We note that in the previous works \cite{Patri2020b, Patri2020c}, there was a minor typographical error in the stated definition of the pseudospin operators (missing the above 1/2 factor of Eq. A9); nevertheless, the results from those works \cite{Patri2020b, Patri2020c} employ the correct definition of the pseudospin operators given in Eq. A9 (i.e. with the factor of 1/2).

\section{\label{app:symmetries} Action of Tetrahedral Group}
In order to test which terms in the Hamiltonian are allowed, we need to know how candidate terms transform under action of the tetrahedral group $T_d$, and under time-reversal $\mathcal{T}$. The most economical way to check all transformations is pick two generators of $T_d$, which are $\mathcal{C}_{31}$ and $\mathcal{S}_{4z}$.

\begin{table}[ht]
\begin{tabular}{c|ccc}
Object & $\mathcal{T}$ & $\mathcal{C}_{31}$ & $\mathcal{S}_{4z}$ \\ \hline
$x$ & $x$ & $z$ & $y$ \\
$y$ & $y$ & $x$ & $-x$ \\
$z$ & $z$ & $y$ & $-z$ \\
$\hat{S}^x$ & $\hat{S}^x$ & $-\frac{1}{2}\hat{S}^x + \frac{\sqrt{3}}{2}\hat{S}^y$ & $-\hat{S}^x$ \\
$\hat{S}^y$ & $\hat{S}^y$ & $-\frac{\sqrt{3}}{2}\hat{S}^x - \frac{1}{2}\hat{S}^y$ & $\hat{S}^y$ \\
$\hat{S}^z$ & $-\hat{S}^z$ & $\hat{S}^z$ & $-\hat{S}^z$ \\
$x(y^2-z^2)$ & $x(y^2-z^2)$ & $ z(x^2-y^2)$ & $ -y(z^2-x^2)$ \\
$y(z^2-x^2)$ & $y(z^2-x^2)$ & $ x(y^2-z^2)$ & $ x(y^2-z^2)$ \\
$z(x^2-y^2)$ & $z(x^2-y^2)$ & $ y(z^2-x^2)$ & $ z(x^2-y^2)$ \\
$\sigma^x$ & $-\sigma^x$ & $\sigma^z$ & $-\sigma^y$ \\
$\sigma^y$ & $-\sigma^y$ & $\sigma^x$ & $\sigma^x$ \\
$\sigma^z$ & $-\sigma^z$ & $\sigma^y$ & $\sigma^z$ 
\end{tabular}
\caption{Symmetry transformations of various objects under two generators of the tetrahedral group.}
\end{table}

\noindent The 11 Kondo Hamiltonians are constructed to be the most general ones which respect time-reversal symmetry and the tetrahedral symmetry. 

\section{ \label{app:basis_change} Cubic Basis to Composite Spin Basis}
In order to express the Hamiltonian with respect to the composite spin basis, we first express the cubic harmonics in terms of spherical harmonics. The three $T_1$ and three $T_2$ orbitals are expressed in the following linear combinations,

\begin{align}
\ket{x(y^2-z^2)} =& \frac{\sqrt{5}}{4}\left(\ket{3,1} - \ket{3,-1}\right) + \frac{\sqrt{3}}{4}\left( \ket{3,3} -\ket{3,-3}\right), \\
\ket{y(z^2-x^2)} =& \frac{i\sqrt{5}}{4}\left(\ket{3,1} + \ket{3,-1}\right) - \frac{i\sqrt{3}}{4}\left(\ket{3,3} +\ket{3,-3}\right), \\
\ket{z(x^2-y^2)} =& \frac{\sqrt{2}}{2}\left(\ket{3,2} + \ket{3,-2}\right), \\
\ket{x} =& \frac{\sqrt{2}}{2}\left(-\ket{1,1} + \ket{1,-1}\right), \\
\ket{y} =& \frac{i\sqrt{2}}{2}\left(\ket{1,1} + \ket{1,-1}\right), \\
\ket{z} =& \ket{1,0}.
\end{align}

After doing this, we look at the total electronic state $\ket{\text{orbital}}\otimes\ket{\text{spin}}$ and use the Clebsch-Gordan coefficients to rewrite them in terms of the composite spin basis. The Clebsch-Gordan transformation is enumerated in the following equations:

\begin{align}
\ket{3,m}\ket{\pm\frac{1}{2}} = \sqrt{\frac{4\pm m}{7}} \ket{\frac{7}{2}, m\pm \frac{1}{2}} \mp  \sqrt{\frac{3\mp m}{7}}\ket{\frac{5}{2}, m\pm \frac{1}{2}}, \\
\ket{1,m}\ket{\pm\frac{1}{2}} = \sqrt{\frac{2\pm m}{3}} \ket{\frac{3}{2}, m\pm \frac{1}{2}} \mp \sqrt{\frac{1\mp m}{3}}\ket{\frac{1}{2}, m\pm \frac{1}{2}}.
\end{align}

\section{Multipolar Kondo Models}
To construct the multipolar Kondo interaction, we first start by writing the Hamiltonian in terms of possible transitions between cubic harmonic states corresponding to the basis functions of $T_1$ and $T_2$. Although this is very simple to do symmetry analysis on, the physical structure of this is not very illuminating. To uncover the physics of the Hamiltonian, it is advantageous to first take the cubic harmonics and write them in terms of spherical harmonics, and then take the resulting composite spin system and use the Clebsch-Gordan coefficients to express in terms of $\ket{j',m_{j'}}\bra{j,m_j}$ operators using the procedure in \ref{app:basis_change}. 

\subsection{\label{app:T1_model} $T_1$-orbital Kondo Model}

As mentioned in the main text, transforming from cubic basis to SO-coupled basis preserves the number of conduction electrons i.e. takes $3\times 2 = 6$ conduction electron states for 3 orbital states and a spin up/down state.
This, however, appears to artificially introduce states to make $(2\times \frac{5}{2} + 1) + (2\times \frac{7}{2} + 1) = 14$ states. This is in fact not the case because only certain linear combinations of the composite-spin kets appear in the Hamiltonian. We label these special linear combinations (Eqs. \ref{eq:chi1}, \ref{eq:chi2}, \ref{eq:chi3} in the main text) by $\ket{\chi^{(\pm)}_i}$, where $i=1,2,3$, and the $(\pm)$ superscript denotes two time-reversal related pairs of these special linear combinations. It should also be noted that this transformation is unitary.
This brings us back to the 6 possible states, which one naturally expects.

To write down the Hamiltonian for the $T_1$ orbitals, we find the symmetry-allowed terms according to $T_d$ symmetry and time-reversal symmetry. We use the standard $3\times 3$ Gell-Mann matrices $\lambda^k$ to express linear combinations of conduction orbitals. The use of these matrices simultaneously guarantees self-adjointness. The conduction electron's spin is described with the standard $\sigma$ Pauli matrices. We express the symmetry-allowed Hamiltonians in Eqs. \eqref{eq:kq1_coupling} - \eqref{eq:ko_coupling} in terms of three coupling constants $K_{Q1}$, $K_{Q2}$, and $K_O$. Note that these are not the same couplings as listed in Section \ref{sec:f_model}. After the change of basis, there is a more natural way to group the operators, which calls for a redefinition of the constants. The relationship between the two sets is $F_{Q1} = K_{Q1} - \frac{K_{Q2}}{\sqrt{3}}$, $F_{Q2} = \frac{K_{Q1}}{\sqrt{2}} + \sqrt{\frac{2}{3}}K_{Q2}$, and $F_O = \sqrt{3}K_O$. These $F$ constants also greatly simplify the forms of the $\beta$-functions.

\begin{align}H_{Q1} =& K_{Q1}\c^\dagger_{0a\alpha}\left(\sigma^z_{\alpha\beta}\lambda^2_{ab} \hat{S}^y + \sigma^y_{\alpha\beta}\lambda^5_{ab}\left(\frac{\sqrt{3}}{2}\hat{S}^x + \frac{1}{2}\hat{S}^y\right) \right. \nonumber \\
&\left. + \sigma^x_{\alpha\beta}\lambda^7_{ab}\left(\frac{\sqrt{3}}{2}\hat{S}^x - \frac{1}{2}\hat{S}^y\right) \right) \c_{0b\beta}, \label{eq:kq1_coupling}  \\
H_{Q2} =& K_{Q2}\c^\dagger_{0a\alpha}\left(\sigma^0_{\alpha\beta} \lambda^3_{ab} \hat{S}^x - \sigma^0_{\alpha\beta} \lambda^8_{ab} \hat{S}^y \right) \c_{0b\beta},  \label{eq:kq2_coupling} \\
 H_O =& K_O \c^\dagger_{0 a\alpha}\left(\sigma^x_{\alpha\beta}\lambda^6_{ab} \hat{S}^z + \sigma^y_{\alpha\beta} \lambda^4_{ab} \hat{S}^z + \sigma^z_{\alpha\beta}\lambda^1_{ab} \hat{S}^z \right) \c_{0b\beta}. \label{eq:ko_coupling} 
\end{align}

The parameters $a,b$ run over orbital wave functions with $T_1$ symmetry. $a=1,2,3$ corresponds to $\ket{x(y^2-z^2)}, \ket{y(z^2-x^2)}$, and $\ket{z(x^2-y^2)}$ respectively. The parameters $\alpha,\beta$ run over the conduction electron's spin, either $\ket{\uparrow}$ or $\ket{\downarrow}$ for $\alpha=1,2$ respectively. The subscript $0$ denotes that these creation/annihilation operators are at the origin (impurity site).

\subsection{\label{app:T2_model} $T_2$-orbital Kondo Model}
The $T_2$ model, which is isomorphic to the $T_1$-orbital model, has been derived and studied in previous work \cite{Patri2020b}. The version after change of basis was used in another previous work \cite{Patri2020c}, and this form is quoted here.

\begin{widetext}
\begin{align}
H^{T_2}_{Q1} =& P_{Q1} \left[ \hat{S}^x\left\{\ket{\frac{3}{2},\frac{1}{2}}\bra{\frac{3}{2},\frac{-3}{2}} + \ket{\frac{3}{2},\frac{3}{2}}\bra{\frac{3}{2},\frac{-1}{2}} + \hc\right\}  \right.  \nonumber \\
& +\left. \hat{S}^y\left\{\ket{\frac{3}{2},\frac{-3}{2}}\bra{\frac{3}{2},\frac{-3}{2}} + \ket{\frac{3}{2},\frac{3}{2}}\bra{\frac{3}{2},\frac{3}{2}} - \ket{\frac{3}{2},\frac{1}{2}}\bra{\frac{3}{2},\frac{1}{2}} - \ket{\frac{3}{2},\frac{-1}{2}}\bra{\frac{3}{2},\frac{-1}{2}}\right\}\right] \label{Hp1}\\
H^{T_2}_{Q2} =& P_{Q2} \left[\hat{S}^x\left\{\ket{\frac{3}{2},\frac{3}{2}}\bra{\frac{1}{2},\frac{-1}{2}} - \ket{\frac{3}{2},\frac{-3}{2}}\bra{\frac{1}{2},\frac{1}{2}}\right\} \right. \nonumber \\
&+\left. \hat{S}^y\left\{\ket{\frac{3}{2},\frac{-1}{2}}\bra{\frac{1}{2},\frac{-1}{2}} - \ket{\frac{3}{2},\frac{1}{2}}\bra{\frac{1}{2},\frac{1}{2}}\right\} + \hc \right] \label{Hp2} \\
H^{T_2}_O=& P_O \hat{S}^z\left[i\ket{\frac{3}{2},\frac{-3}{2}}\bra{\frac{3}{2},\frac{1}{2}} + i\ket{\frac{3}{2},\frac{3}{2}}\bra{\frac{3}{2},\frac{-1}{2}} + \hc \right] \label{Hp3}
\end{align}
\end{widetext}

\subsection{$T_1\otimes T_2$-orbital Kondo Model}
We now use all 12 states to construct possible transitions between states in the different irreducible representations. This leads to the interaction terms in the main text: Eqs. \eqref{eq:Hfp1} - \eqref{eq:Hfp5}.

\section{Many-Body Perturbation Theory}
\label{app_feynman}
When doing perturbation theory, we need to expand $n$-point Green functions, which requires suitable commutation relations on the operators. Since spin operators satisfy a $\algsu(2)$ algebra instead of a canonical anticommutation one, we need to introduce Abrikosov pseudofermions via the relation in \eqref{eq:pseudofermion}, with $\sigma^i$ the $i$-th Pauli matrix.

\begin{equation}
\hat{S}^i = \frac{1}{2} \sum_{\mu,\mu' = \uparrow,\downarrow} \hat{f}^\dagger_{\mu'}\sigma^i_{\mu'\mu} \hat{f}_\mu \label{eq:pseudofermion}
\end{equation}

This artificially allows the pseudofermion to change its occupation number, instead of simply changing its state while remaining at occupation 1. To remedy this, we apply the Popov-Fedotov trick by introducing a complex chemical potential $\lambda = \frac{i\pi}{2\beta}$ for the pseudofermions, which exactly cancels the empty and doubly occupied states from the partition function. In order to calculate the effective vertices at third order perturbation theory, we sum the 2-particle irreducible connected Feynman diagrams $\Gamma_\text{irr, conn}$. Note that 2-particle reducible connected diagrams technically enter the perturbation series, but they only impact the kinetic term, not the vertices. 

\begin{figure}[ht]
\includegraphics{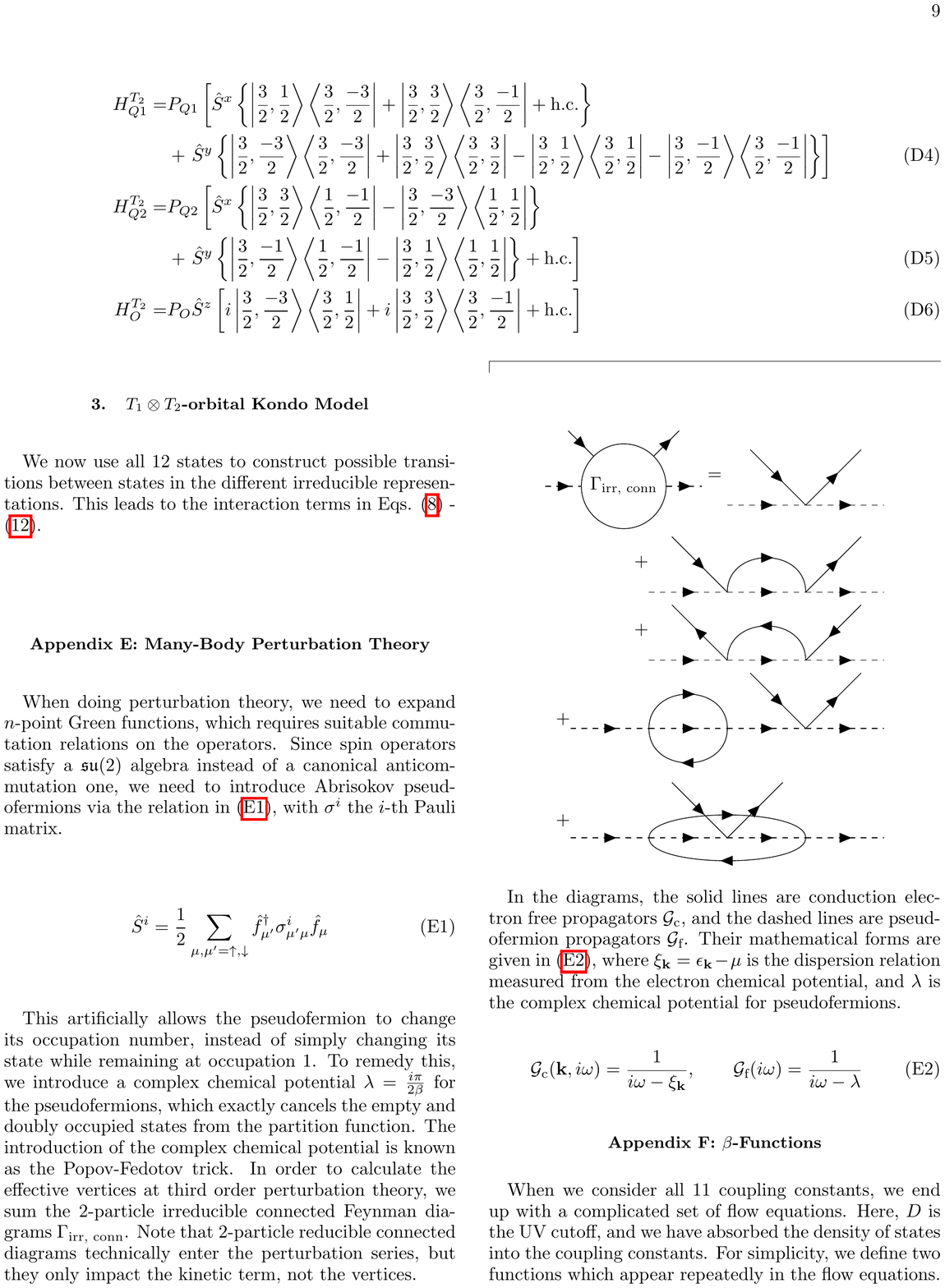}
\caption{Irreducible, connected Feynman diagrams to 2-loop order}
\end{figure}

In the diagrams, the solid lines are conduction electron free propagators $\mathcal{G}_\text{c}$, and the dashed lines are pseudofermion propagators $\mathcal{G}_\text{f}$. Their mathematical forms are given in \eqref{eq:propagators}, where $\xi_\k = \epsilon_\k - \mu$ is the dispersion relation measured from the electron chemical potential, and $\lambda$ is the complex chemical potential for pseudofermions. 

\begin{equation}
\mathcal{G}_\text{c}(\k,i\omega) = \frac{1}{i\omega - \xi_\k} ,\qquad \mathcal{G}_\text{f}(i\omega) = \frac{1}{i\omega - \lambda} \label{eq:propagators} 
\end{equation}

\section{$\beta$-Functions}
When we consider all 11 coupling constants, we end up with a complicated set of flow equations. Here, $D$ is the UV cutoff, and we have absorbed the density of states into the coupling constants. For simplicity, we define two functions which appear repeatedly in the flow equations. 

\begin{align}
W_Q =& 2\left(F_{O}^{2} + F_{Q1}^{2} + F_{Q2}^{2} + P_{O}^{2} + P_{Q1}^{2} + P_{Q2}^{2} \right. \nonumber \\
& \left. + X_{O1}^{2} + 2 X_{O2}^{2} + 2 X_{Q1}^{2} + X_{Q2}^{2} + X_{Q3}^{2}\right) \\
W_O =&  4\left(F_{Q1}^{2} + F_{Q2}^{2} + P_{Q1}^{2} + P_{Q2}^{2} + 2 X_{Q1}^{2} + X_{Q2}^{2} + X_{Q3}^{2}\right)
\end{align}

\begin{align}
\frac{\d F_{Q1}}{\d\log D} =& - 2 F_{O} F_{Q1} - 2 X_{O2} X_{Q1} + F_{Q1} W_Q \\
\frac{\d F_{Q2}}{\d\log D} =& F_{O} F_{Q2} + X_{O1} X_{Q2} + X_{O2} X_{Q3} + F_{Q2} W_Q \\
\frac{\d F_O}{\d\log D} =& - 2 F_{Q1}^{2} + F_{Q2}^{2} - 2 X_{Q1}^{2} + X_{Q2}^{2} + F_{O} W_O
\end{align}

\begin{align}
\frac{\d P_{Q1}}{\d\log D} =& - 2 P_{O} P_{Q1} + 2 X_{O2} X_{Q1} + P_{Q1}W_Q \\
\frac{\d P_{Q2}}{\d\log D} =& P_{O} P_{Q2} - X_{O1} X_{Q3} - X_{O2} X_{Q2} + P_{Q2}W_Q \\
\frac{\d P_O}{\d\log D} =& - 2 P_{Q1}^{2} + P_{Q2}^{2} - 2 X_{Q1}^{2} + X_{Q3}^{2} + P_{O}W_O 
\end{align}

\begin{widetext}
\begin{align}
\frac{\d X_{Q1}}{\d\log D} =& - F_{O} X_{Q1} - F_{Q1} X_{O2} - P_{O} X_{Q1} + P_{Q1} X_{O2} + X_{Q1} W_Q \\
\frac{\d X_{Q2}}{\d\log D} =& F_{O} X_{Q2} + F_{Q2} X_{O1} - P_{Q2} X_{O2} + X_{Q2}W_Q\\
\frac{\d X_{Q3}}{\d\log D} =& F_{Q2} X_{O2} + P_{O} X_{Q3} - P_{Q2} X_{O1} + X_{Q3}W_Q \\
\frac{\d X_{O1}}{\d\log D} =& 2 F_{Q2} X_{Q2} - 2 P_{Q2} X_{Q3} + X_{O1}W_O\\
\frac{\d X_{O2}}{\d\log D} =& - 2 F_{Q1} X_{Q1} + F_{Q2} X_{Q3} + 2 P_{Q1} X_{Q1} - P_{Q2} X_{Q2} + X_{O2}W_O
\end{align}
\end{widetext}

\section{ \label{app:fixed_points} Fixed Points}
We find two stable fixed manifolds to the flow equations. First, the two-channel fixed manifold $M$ has two distinct branches, whose equations are

\begin{align}
& X_{Q2} = 0, \quad X_{Q3} = 0 ,\quad X_{O1} = 0, \quad F_{Q2} = 0,\\
& P_{Q2} = 0, \quad F_O\in\left[0,\frac{1}{2}\right], \quad P_O = \frac{1}{2} - F_O, \nonumber \\
& X_{Q1} = \pm\sqrt{P_OF_O},\quad X_{O2} = \mp \sqrt{P_OF_O}, \nonumber \\
& F_{Q1} = - F_O, \quad P_{Q1} = P_O, \nonumber 
\end{align}

and there is another branch with 

\begin{align}
& X_{Q2} = 0, \quad X_{Q3} = 0 ,\quad X_{O1} = 0, \quad F_{Q2} = 0, \\
& P_{Q2} = 0, \quad F_O\in\left[0,\frac{1}{2}\right],\quad P_O = \frac{1}{2} - F_O, \nonumber \\
& X_{Q1} = \pm\sqrt{P_OF_O}, \quad X_{O2} = \mp \sqrt{P_OF_O}, \nonumber \\
& F_{Q1} = F_O, \qquad P_{Q1} = -P_O. \nonumber 
\end{align}

Secondly, the novel fixed manifold $L$ again has two distinct branches, whose equations are 

\begin{align}
& X_{Q1} = 0, \quad X_{O1} = \frac{1}{4}, \quad F_{Q1} = 0,\quad P_{Q1} = 0,  \\
& F_O \in \left[-\frac{1}{4},0\right],\quad P_O = -\frac{1}{4} - F_O, \quad X_{O2} = -\sqrt{F_OP_O}, \nonumber \\
& X_{Q3} = \pm\frac{P_O - F_O - 2\sqrt{P_OF_O} - 1/4}{2\sqrt{1 + 8\sqrt{F_OP_O}}}, \nonumber \\
& X_{Q2} = \pm\frac{P_O - F_O + 2\sqrt{P_OF_O} + 1/4}{2\sqrt{1 + 8\sqrt{F_OP_O}}}, \nonumber \\
& F_{Q2} = \pm \frac{F_O - \sqrt{F_OP_O}}{\sqrt{1+8\sqrt{F_OP_O}}}, \quad P_{Q2} = \pm \frac{P_O - \sqrt{F_OP_O}}{\sqrt{1+8\sqrt{F_OP_O}}}, \nonumber 
\end{align}

and there is another branch with 

\begin{align}
& X_{Q1} = 0, \quad X_{O1} = -\frac{1}{4},\quad F_{Q1} = 0,\quad P_{Q1} = 0, \\ 
& F_O \in \left[-\frac{1}{4},0\right],\quad P_O = -\frac{1}{4} - F_O,\quad X_{O2} = \sqrt{F_OP_O}, \nonumber \\ 
& X_{Q3} = \pm\frac{P_O - F_O - 2\sqrt{P_OF_O} - 1/4}{2\sqrt{1 + 8\sqrt{F_OP_O}}}, \nonumber \\ 
& X_{Q2} = \pm\frac{P_O - F_O + 2\sqrt{P_OF_O} + 1/4}{2\sqrt{1 + 8\sqrt{F_OP_O}}}, \nonumber \\ 
& F_{Q2} = \mp\frac{F_O - \sqrt{P_OF_O}}{\sqrt{1+8\sqrt{F_OP_O}}},\quad P_{Q2} = \mp \frac{P_O - \sqrt{P_OF_O}}{\sqrt{1+8\sqrt{P_OF_O}}}. \nonumber 
\end{align}

We note that the scaling behavior $\Delta$ is the same at any point on a particular manifold.

\section{Unitary transformation mapping Kondo model to four-channel Kondo model}
\label{app:diagonal_basis}

As described in the main text, a series of unitary transformations is performed that maps the discovered fixed point manifold ($L$) to the four-channel Kondo model.
For clarity, the unitary transformation that diagonalizes $\left(T^2 + T^{14} \right)$ in each channel $s$ yields eigenstates,
\begin{align}
& \hat{\psi}_{s, 1} = \frac{1}{\sqrt{2}} \left(  i \c_{s,1} + \c_{s,2}\right), \\ 
& \hat{\psi}_{s, 2} = \frac{1}{\sqrt{2}} \left(  i \c_{s,3} + \c_{s,4}\right), \\
& \hat{\psi}_{s, 3} = \frac{1}{\sqrt{2}} \left(  -i \c_{s,1} + \c_{s,2}\right), \\
& \hat{\psi}_{s, 4} = \frac{1}{\sqrt{2}} \left(  -i \c_{s,3} + \c_{s,4}\right).
\end{align}

\section{Strong Coupling Analysis}
In order to establish the validity and existence of the intermediate fixed point, it is important to verify that the corresponding strong coupling limit is unstable. In the strong coupling limit, we work with the Hamiltonian at the fixed point in a single coupling constant, Eq. \eqref{eq:L_point_model}. Taking $g\to\infty$ eliminates the kinetic energy term to form a single-site system. This single site problem can be solved exactly to yield a 4-fold degenerate ground state. The scattering within the 4-fold degenerate manifold destabilizes the strong coupling fixed point which requires flow back into the perturbative window \cite{Nozieres1980}.

These 4 states, in Eqs. \eqref{eq:GS1} - \eqref{eq:GS4}, are found for the choice $P_{Q2} = P_O = -X_{Q3} = X_{O1} = -1/4$. There are alternative choices which will lead to the same degeneracy and eigenvalues, but the eigenvectors will differ slightly. Here, every operator is on the impurity site, and the $\ket{\uparrow},\ket{\downarrow}$ kets correspond to the state of the pseudospin. This is not to be confused with the notation in Appendix \ref{app:basis_change}, where the tensor product was between the two different degrees of freedom for the conduction electron; here the tensor product is between the conduction electron state in the composite spin-orbit coupled basis and the pseudospin.

\begin{widetext}
\begin{align}
\ket{\psi_1} =& \frac{1}{2}\left(i\ket{\frac{1}{2},\frac{1}{2}} - \ket{\chi^{(+)}_3}\right)\otimes\ket{\uparrow} + \frac{1}{2}\left(-i\ket{\frac{3}{2},\frac{-3}{2}} + \ket{\frac{3}{2},\frac{1}{2}}\right)\otimes\ket{\downarrow} \label{eq:GS1}  \\
\ket{\psi_2} =& \frac{1}{2}\left(-i\ket{\frac{3}{2},\frac{-3}{2}} - \ket{\frac{3}{2},\frac{1}{2}}\right)\otimes\ket{\uparrow} + \frac{1}{2}\left(i\ket{\frac{1}{2},\frac{1}{2}} + \ket{\chi^{(+)}_3}\right)\otimes\ket{\downarrow} \label{eq:GS2} \\
\ket{\psi_3} =& \frac{1}{2}\left(\ket{\frac{1}{2},\frac{-1}{2}} + i\ket{\chi^{(-)}_3}\right)\otimes\ket{\uparrow} + \frac{1}{2}\left(i\ket{\frac{3}{2},\frac{-1}{2}} + \ket{\frac{3}{2},\frac{3}{2}}\right)\otimes\ket{\downarrow} \label{eq:GS3} \\
\ket{\psi_4} =& \frac{1}{2}\left(\ket{\frac{3}{2},\frac{-1}{2}} + i\ket{\frac{3}{2},\frac{3}{2}}\right)\otimes\ket{\downarrow} + \frac{1}{2}\left(i\ket{\frac{1}{2},\frac{-1}{2}} + \ket{\chi^{(-)}_3}\right)\otimes\ket{\downarrow} \label{eq:GS4}
\end{align}
\end{widetext}

The next step in the strong coupling analysis is to consider coupling of the impurity to the conduction electrons. The impurity is at site $0$, and the two additional sites to which conduction electrons can hop are at $\hat{z}$ and $-\hat{z}$. We perform standard Slater-Koster tight-binding for this kinetic term.
We note that not every transition is possible between the 6 possible orbitals (3 $T_1$ orbitals and 3 $T_2$ orbitals) on each site, and some hopping weights must be equal by symmetry. The Slater-Koster analysis yields the $6\times 6$ hopping matrix $T$, with . The Hamiltonian for the system is thus 

\begin{equation}
\hat{H} = -\sum_{ab\sigma} T_{ab}(\c^\dagger_{\hat{z},a,\sigma}  \c_{0b,\sigma} + \c^\dagger_{-\hat{z},a,\sigma} \c_{0,b,\sigma} + \hc) + \hat{H}_{L^*}
\end{equation}
where $H_{L^*}$ is same interaction Hamiltonian from Eq. \eqref{eq:L_point_model}. The 4-fold degeneracy of the single-site problem is split into two 2-fold degenerate doublets when this hopping is added. The 2-fold degeneracy of the new ground state confirms the instability of the strong coupling limit, and that the renormalization group flow returns back to the discovered intermediate fixed point.

\end{appendix}


\begin{thebibliography}{40}%
\makeatletter
\providecommand \@ifxundefined [1]{%
 \@ifx{#1\undefined}
}%
\providecommand \@ifnum [1]{%
 \ifnum #1\expandafter \@firstoftwo
 \else \expandafter \@secondoftwo
 \fi
}%
\providecommand \@ifx [1]{%
 \ifx #1\expandafter \@firstoftwo
 \else \expandafter \@secondoftwo
 \fi
}%
\providecommand \natexlab [1]{#1}%
\providecommand \enquote  [1]{``#1''}%
\providecommand \bibnamefont  [1]{#1}%
\providecommand \bibfnamefont [1]{#1}%
\providecommand \citenamefont [1]{#1}%
\providecommand \href@noop [0]{\@secondoftwo}%
\providecommand \href [0]{\begingroup \@sanitize@url \@href}%
\providecommand \@href[1]{\@@startlink{#1}\@@href}%
\providecommand \@@href[1]{\endgroup#1\@@endlink}%
\providecommand \@sanitize@url [0]{\catcode `\\12\catcode `\$12\catcode
  `\&12\catcode `\#12\catcode `\^12\catcode `\_12\catcode `\%12\relax}%
\providecommand \@@startlink[1]{}%
\providecommand \@@endlink[0]{}%
\providecommand \url  [0]{\begingroup\@sanitize@url \@url }%
\providecommand \@url [1]{\endgroup\@href {#1}{\urlprefix }}%
\providecommand \urlprefix  [0]{URL }%
\providecommand \Eprint [0]{\href }%
\providecommand \doibase [0]{http://dx.doi.org/}%
\providecommand \selectlanguage [0]{\@gobble}%
\providecommand \bibinfo  [0]{\@secondoftwo}%
\providecommand \bibfield  [0]{\@secondoftwo}%
\providecommand \translation [1]{[#1]}%
\providecommand \BibitemOpen [0]{}%
\providecommand \bibitemStop [0]{}%
\providecommand \bibitemNoStop [0]{.\EOS\space}%
\providecommand \EOS [0]{\spacefactor3000\relax}%
\providecommand \BibitemShut  [1]{\csname bibitem#1\endcsname}%
\let\auto@bib@innerbib\@empty
\bibitem [{\citenamefont {Keimer}\ \emph {et~al.}(2015)\citenamefont {Keimer},
  \citenamefont {Kivelson}, \citenamefont {Norman}, \citenamefont {Uchida},\
  and\ \citenamefont {Zaanen}}]{Keimer2015a}%
  \BibitemOpen
  \bibfield  {author} {\bibinfo {author} {\bibfnamefont {B.}~\bibnamefont
  {Keimer}}, \bibinfo {author} {\bibfnamefont {S.~A.}\ \bibnamefont
  {Kivelson}}, \bibinfo {author} {\bibfnamefont {M.~R.}\ \bibnamefont
  {Norman}}, \bibinfo {author} {\bibfnamefont {S.}~\bibnamefont {Uchida}}, \
  and\ \bibinfo {author} {\bibfnamefont {J.}~\bibnamefont {Zaanen}},\
  }\bibfield  {title} {\enquote {\bibinfo {title} {{From quantum matter to
  high-temperature superconductivity in copper oxides}},}\ }\href {\doibase
  10.1038/nature14165} {\bibfield  {journal} {\bibinfo  {journal} {Nature}\
  }\textbf {\bibinfo {volume} {518}},\ \bibinfo {pages} {179--186} (\bibinfo
  {year} {2015})}\BibitemShut {NoStop}%
\bibitem [{\citenamefont {Si}\ \emph {et~al.}(2016)\citenamefont {Si},
  \citenamefont {Yu},\ and\ \citenamefont {Abrahams}}]{Si2016}%
  \BibitemOpen
  \bibfield  {author} {\bibinfo {author} {\bibfnamefont {Qimiao}\ \bibnamefont
  {Si}}, \bibinfo {author} {\bibfnamefont {Rong}\ \bibnamefont {Yu}}, \ and\
  \bibinfo {author} {\bibfnamefont {Elihu}\ \bibnamefont {Abrahams}},\
  }\bibfield  {title} {\enquote {\bibinfo {title} {{High-temperature
  superconductivity in iron pnictides and chalcogenides}},}\ }\href {\doibase
  10.1038/natrevmats.2016.17} {\bibfield  {journal} {\bibinfo  {journal}
  {Nature Reviews Materials}\ }\textbf {\bibinfo {volume} {1}} (\bibinfo {year}
  {2016}),\ 10.1038/natrevmats.2016.17}\BibitemShut {NoStop}%
\bibitem [{\citenamefont {Chubukov}(2012)}]{Chubukov2012}%
  \BibitemOpen
  \bibfield  {author} {\bibinfo {author} {\bibfnamefont {Andrey}\ \bibnamefont
  {Chubukov}},\ }\bibfield  {title} {\enquote {\bibinfo {title} {{Pairing
  mechanism in fe-based superconductors}},}\ }\href {\doibase
  10.1146/annurev-conmatphys-020911-125055} {\bibfield  {journal} {\bibinfo
  {journal} {Annual Review of Condensed Matter Physics}\ }\textbf {\bibinfo
  {volume} {3}},\ \bibinfo {pages} {57--92} (\bibinfo {year}
  {2012})}\BibitemShut {NoStop}%
\bibitem [{\citenamefont {Coleman}(2017)}]{Coleman2017}%
  \BibitemOpen
  \bibfield  {author} {\bibinfo {author} {\bibfnamefont {Piers}\ \bibnamefont
  {Coleman}},\ }\bibfield  {title} {\enquote {\bibinfo {title} {{Theory
  perspective: SCES 2016}},}\ }\href {\doibase 10.1080/14786435.2016.1232866}
  {\bibfield  {journal} {\bibinfo  {journal} {Philosophical Magazine}\ }\textbf
  {\bibinfo {volume} {97}},\ \bibinfo {pages} {3527--3543} (\bibinfo {year}
  {2017})}\BibitemShut {NoStop}%
\bibitem [{\citenamefont {Lee}(2018)}]{Lee2018a}%
  \BibitemOpen
  \bibfield  {author} {\bibinfo {author} {\bibfnamefont {Sung~Sik}\
  \bibnamefont {Lee}},\ }\bibfield  {title} {\enquote {\bibinfo {title}
  {{Recent Developments in Non-Fermi Liquid Theory}},}\ }\href {\doibase
  10.1146/annurev-conmatphys-031016-025531} {\bibfield  {journal} {\bibinfo
  {journal} {Annual Review of Condensed Matter Physics}\ }\textbf {\bibinfo
  {volume} {9}},\ \bibinfo {pages} {227--244} (\bibinfo {year}
  {2018})}\BibitemShut {NoStop}%
\bibitem [{\citenamefont {Stewart}(2001)}]{Stewart2001}%
  \BibitemOpen
  \bibfield  {author} {\bibinfo {author} {\bibfnamefont {G.~R.}\ \bibnamefont
  {Stewart}},\ }\bibfield  {title} {\enquote {\bibinfo {title}
  {{Non-Fermi-liquid behavior in d- and f-electron metals}},}\ }\href {\doibase
  10.1103/RevModPhys.73.797} {\bibfield  {journal} {\bibinfo  {journal}
  {Reviews of Modern Physics}\ }\textbf {\bibinfo {volume} {73}},\ \bibinfo
  {pages} {797--855} (\bibinfo {year} {2001})}\BibitemShut {NoStop}%
\bibitem [{\citenamefont {Cao}\ \emph {et~al.}(2020)\citenamefont {Cao},
  \citenamefont {Chowdhury}, \citenamefont {Rodan-Legrain}, \citenamefont
  {Rubies-Bigorda}, \citenamefont {Watanabe}, \citenamefont {Taniguchi},
  \citenamefont {Senthil},\ and\ \citenamefont {Jarillo-Herrero}}]{Cao2020}%
  \BibitemOpen
  \bibfield  {author} {\bibinfo {author} {\bibfnamefont {Yuan}\ \bibnamefont
  {Cao}}, \bibinfo {author} {\bibfnamefont {Debanjan}\ \bibnamefont
  {Chowdhury}}, \bibinfo {author} {\bibfnamefont {Daniel}\ \bibnamefont
  {Rodan-Legrain}}, \bibinfo {author} {\bibfnamefont {Oriol}\ \bibnamefont
  {Rubies-Bigorda}}, \bibinfo {author} {\bibfnamefont {Kenji}\ \bibnamefont
  {Watanabe}}, \bibinfo {author} {\bibfnamefont {Takashi}\ \bibnamefont
  {Taniguchi}}, \bibinfo {author} {\bibfnamefont {T.}~\bibnamefont {Senthil}},
  \ and\ \bibinfo {author} {\bibfnamefont {Pablo}\ \bibnamefont
  {Jarillo-Herrero}},\ }\bibfield  {title} {\enquote {\bibinfo {title}
  {{Strange Metal in Magic-Angle Graphene with near Planckian Dissipation}},}\
  }\href {\doibase 10.1103/PhysRevLett.124.076801} {\bibfield  {journal}
  {\bibinfo  {journal} {Physical Review Letters}\ }\textbf {\bibinfo {volume}
  {124}},\ \bibinfo {pages} {076801} (\bibinfo {year} {2020})}\BibitemShut
  {NoStop}%
\bibitem [{\citenamefont {Ludwig}(1994)}]{Ludwig1994}%
  \BibitemOpen
  \bibfield  {author} {\bibinfo {author} {\bibfnamefont {Andreas~W.W.}\
  \bibnamefont {Ludwig}},\ }\bibfield  {title} {\enquote {\bibinfo {title}
  {{Exact results on the multi-channel Kondo effect from conformal field
  theory}},}\ }\href {\doibase 10.1016/0921-4526(94)91851-1} {\bibfield
  {journal} {\bibinfo  {journal} {Physica B: Physics of Condensed Matter}\
  }\textbf {\bibinfo {volume} {199-200}},\ \bibinfo {pages} {406--408}
  (\bibinfo {year} {1994})}\BibitemShut {NoStop}%
\bibitem [{\citenamefont {Andrei}\ \emph {et~al.}(1983)\citenamefont {Andrei},
  \citenamefont {Furuya},\ and\ \citenamefont {Lowenstein}}]{Andrei1983}%
  \BibitemOpen
  \bibfield  {author} {\bibinfo {author} {\bibfnamefont {N.}~\bibnamefont
  {Andrei}}, \bibinfo {author} {\bibfnamefont {K.}~\bibnamefont {Furuya}}, \
  and\ \bibinfo {author} {\bibfnamefont {J.~H.}\ \bibnamefont {Lowenstein}},\
  }\bibfield  {title} {\enquote {\bibinfo {title} {{Solution of the Kondo
  problem}},}\ }\href {\doibase 10.1103/RevModPhys.55.331} {\bibfield
  {journal} {\bibinfo  {journal} {Reviews of Modern Physics}\ }\textbf
  {\bibinfo {volume} {55}},\ \bibinfo {pages} {331--402} (\bibinfo {year}
  {1983})}\BibitemShut {NoStop}%
\bibitem [{\citenamefont {Tsvelick}\ and\ \citenamefont
  {Wiegmann}(1984)}]{Tsvelick1984}%
  \BibitemOpen
  \bibfield  {author} {\bibinfo {author} {\bibfnamefont {A.~M.}\ \bibnamefont
  {Tsvelick}}\ and\ \bibinfo {author} {\bibfnamefont {P.~B.}\ \bibnamefont
  {Wiegmann}},\ }\bibfield  {title} {\enquote {\bibinfo {title} {{Solution of
  the n-channel Kondo problem (scaling and integrability)}},}\ }\href {\doibase
  10.1007/BF01319184} {\bibfield  {journal} {\bibinfo  {journal} {Zeitschrift
  f{\"{u}}r Physik B Condensed Matter}\ }\textbf {\bibinfo {volume} {54}},\
  \bibinfo {pages} {201--206} (\bibinfo {year} {1984})}\BibitemShut {NoStop}%
\bibitem [{\citenamefont {Tsvelick}\ and\ \citenamefont
  {Wiegmann}(1985)}]{Tsvelick1985}%
  \BibitemOpen
  \bibfield  {author} {\bibinfo {author} {\bibfnamefont {A.~M.}\ \bibnamefont
  {Tsvelick}}\ and\ \bibinfo {author} {\bibfnamefont {P.~B.}\ \bibnamefont
  {Wiegmann}},\ }\bibfield  {title} {\enquote {\bibinfo {title} {{Exact
  solution of the multichannel Kondo problem, scaling, and integrability}},}\
  }\href {\doibase 10.1007/BF01017853} {\bibfield  {journal} {\bibinfo
  {journal} {Journal of Statistical Physics}\ }\textbf {\bibinfo {volume}
  {38}},\ \bibinfo {pages} {125--147} (\bibinfo {year} {1985})}\BibitemShut
  {NoStop}%
\bibitem [{\citenamefont {Onimaru}\ \emph {et~al.}(2011)\citenamefont
  {Onimaru}, \citenamefont {Matsumoto}, \citenamefont {Inoue}, \citenamefont
  {Umeo}, \citenamefont {Sakakibara}, \citenamefont {Karaki}, \citenamefont
  {Kubota},\ and\ \citenamefont {Takabatake}}]{Onimaru2011}%
  \BibitemOpen
  \bibfield  {author} {\bibinfo {author} {\bibfnamefont {T.}~\bibnamefont
  {Onimaru}}, \bibinfo {author} {\bibfnamefont {K.~T.}\ \bibnamefont
  {Matsumoto}}, \bibinfo {author} {\bibfnamefont {Y.~F.}\ \bibnamefont
  {Inoue}}, \bibinfo {author} {\bibfnamefont {K.}~\bibnamefont {Umeo}},
  \bibinfo {author} {\bibfnamefont {T.}~\bibnamefont {Sakakibara}}, \bibinfo
  {author} {\bibfnamefont {Y.}~\bibnamefont {Karaki}}, \bibinfo {author}
  {\bibfnamefont {M.}~\bibnamefont {Kubota}}, \ and\ \bibinfo {author}
  {\bibfnamefont {T.}~\bibnamefont {Takabatake}},\ }\bibfield  {title}
  {\enquote {\bibinfo {title} {{Antiferroquadrupolar ordering in a Pr-based
  superconductor PrIr 2Zn20}},}\ }\href {\doibase
  10.1103/PhysRevLett.106.177001} {\bibfield  {journal} {\bibinfo  {journal}
  {Physical Review Letters}\ }\textbf {\bibinfo {volume} {106}},\ \bibinfo
  {pages} {2--5} (\bibinfo {year} {2011})}\BibitemShut {NoStop}%
\bibitem [{\citenamefont {Tsujimoto}\ \emph {et~al.}(2014)\citenamefont
  {Tsujimoto}, \citenamefont {Matsumoto}, \citenamefont {Tomita}, \citenamefont
  {Sakai},\ and\ \citenamefont {Nakatsuji}}]{Tsujimoto2014}%
  \BibitemOpen
  \bibfield  {author} {\bibinfo {author} {\bibfnamefont {Masaki}\ \bibnamefont
  {Tsujimoto}}, \bibinfo {author} {\bibfnamefont {Yosuke}\ \bibnamefont
  {Matsumoto}}, \bibinfo {author} {\bibfnamefont {Takahiro}\ \bibnamefont
  {Tomita}}, \bibinfo {author} {\bibfnamefont {Akito}\ \bibnamefont {Sakai}}, \
  and\ \bibinfo {author} {\bibfnamefont {Satoru}\ \bibnamefont {Nakatsuji}},\
  }\bibfield  {title} {\enquote {\bibinfo {title} {{Heavy-fermion
  superconductivity in the quadrupole ordered state of PrV2Al20}},}\ }\href
  {\doibase 10.1103/PhysRevLett.113.267001} {\bibfield  {journal} {\bibinfo
  {journal} {Physical Review Letters}\ }\textbf {\bibinfo {volume} {113}},\
  \bibinfo {pages} {1--5} (\bibinfo {year} {2014})}\BibitemShut {NoStop}%
\bibitem [{\citenamefont {Sato}\ \emph {et~al.}(2012)\citenamefont {Sato},
  \citenamefont {Ibuka}, \citenamefont {Nambu}, \citenamefont {Yamazaki},
  \citenamefont {Hong}, \citenamefont {Sakai},\ and\ \citenamefont
  {Nakatsuji}}]{Sato2012}%
  \BibitemOpen
  \bibfield  {author} {\bibinfo {author} {\bibfnamefont {Taku~J.}\ \bibnamefont
  {Sato}}, \bibinfo {author} {\bibfnamefont {Soshi}\ \bibnamefont {Ibuka}},
  \bibinfo {author} {\bibfnamefont {Yusuke}\ \bibnamefont {Nambu}}, \bibinfo
  {author} {\bibfnamefont {Teruo}\ \bibnamefont {Yamazaki}}, \bibinfo {author}
  {\bibfnamefont {Tao}\ \bibnamefont {Hong}}, \bibinfo {author} {\bibfnamefont
  {Akito}\ \bibnamefont {Sakai}}, \ and\ \bibinfo {author} {\bibfnamefont
  {Satoru}\ \bibnamefont {Nakatsuji}},\ }\bibfield  {title} {\enquote {\bibinfo
  {title} {{Ferroquadrupolar ordering in PrTi 2Al 20}},}\ }\href {\doibase
  10.1103/PhysRevB.86.184419} {\bibfield  {journal} {\bibinfo  {journal}
  {Physical Review B - Condensed Matter and Materials Physics}\ }\textbf
  {\bibinfo {volume} {86}},\ \bibinfo {pages} {184419} (\bibinfo {year}
  {2012})}\BibitemShut {NoStop}%
\bibitem [{\citenamefont {Sakai}\ \emph {et~al.}(2012)\citenamefont {Sakai},
  \citenamefont {Kuga},\ and\ \citenamefont {Nakatsuji}}]{Sakai2012}%
  \BibitemOpen
  \bibfield  {author} {\bibinfo {author} {\bibfnamefont {Akito}\ \bibnamefont
  {Sakai}}, \bibinfo {author} {\bibfnamefont {Kentaro}\ \bibnamefont {Kuga}}, \
  and\ \bibinfo {author} {\bibfnamefont {Satoru}\ \bibnamefont {Nakatsuji}},\
  }\bibfield  {title} {\enquote {\bibinfo {title} {{Superconductivity in the
  ferroquadrupolar state in the quadrupolar kondo lattice PrTi 2Al 20}},}\
  }\href {\doibase 10.1143/JPSJ.81.083702} {\bibfield  {journal} {\bibinfo
  {journal} {Journal of the Physical Society of Japan}\ }\textbf {\bibinfo
  {volume} {81}},\ \bibinfo {pages} {083702} (\bibinfo {year}
  {2012})}\BibitemShut {NoStop}%
\bibitem [{\citenamefont {Riggs}\ \emph {et~al.}(2015)\citenamefont {Riggs},
  \citenamefont {Shapiro}, \citenamefont {Maharaj}, \citenamefont {Raghu},
  \citenamefont {Bauer}, \citenamefont {Baumbach}, \citenamefont
  {Giraldo-Gallo}, \citenamefont {Wartenbe},\ and\ \citenamefont
  {Fisher}}]{Riggs2015}%
  \BibitemOpen
  \bibfield  {author} {\bibinfo {author} {\bibfnamefont {Scott~C.}\
  \bibnamefont {Riggs}}, \bibinfo {author} {\bibfnamefont {M.C.}\ \bibnamefont
  {Shapiro}}, \bibinfo {author} {\bibfnamefont {Akash~V}\ \bibnamefont
  {Maharaj}}, \bibinfo {author} {\bibfnamefont {S.}~\bibnamefont {Raghu}},
  \bibinfo {author} {\bibfnamefont {E.D.}\ \bibnamefont {Bauer}}, \bibinfo
  {author} {\bibfnamefont {R.E.}\ \bibnamefont {Baumbach}}, \bibinfo {author}
  {\bibfnamefont {P.}~\bibnamefont {Giraldo-Gallo}}, \bibinfo {author}
  {\bibfnamefont {Mark}\ \bibnamefont {Wartenbe}}, \ and\ \bibinfo {author}
  {\bibfnamefont {I.R.}\ \bibnamefont {Fisher}},\ }\bibfield  {title} {\enquote
  {\bibinfo {title} {{Evidence for a nematic component to the hidden-order
  parameter in URu2Si2 from differential elastoresistance measurements}},}\
  }\href {\doibase 10.1038/ncomms7425} {\bibfield  {journal} {\bibinfo
  {journal} {Nature Communications}\ }\textbf {\bibinfo {volume} {6}},\
  \bibinfo {pages} {6425} (\bibinfo {year} {2015})}\BibitemShut {NoStop}%
\bibitem [{\citenamefont {Matsubayashi}\ \emph {et~al.}(2012)\citenamefont
  {Matsubayashi}, \citenamefont {Tanaka}, \citenamefont {Sakai}, \citenamefont
  {Nakatsuji}, \citenamefont {Kubo},\ and\ \citenamefont
  {Uwatoko}}]{Matsubayashi2012b}%
  \BibitemOpen
  \bibfield  {author} {\bibinfo {author} {\bibfnamefont {K.}~\bibnamefont
  {Matsubayashi}}, \bibinfo {author} {\bibfnamefont {T.}~\bibnamefont
  {Tanaka}}, \bibinfo {author} {\bibfnamefont {A.}~\bibnamefont {Sakai}},
  \bibinfo {author} {\bibfnamefont {S.}~\bibnamefont {Nakatsuji}}, \bibinfo
  {author} {\bibfnamefont {Y.}~\bibnamefont {Kubo}}, \ and\ \bibinfo {author}
  {\bibfnamefont {Y.}~\bibnamefont {Uwatoko}},\ }\bibfield  {title} {\enquote
  {\bibinfo {title} {{Pressure-induced heavy fermion superconductivity in the
  nonmagnetic quadrupolar system PrTi 2Al 20}},}\ }\href {\doibase
  10.1103/PhysRevLett.109.187004} {\bibfield  {journal} {\bibinfo  {journal}
  {Physical Review Letters}\ }\textbf {\bibinfo {volume} {109}},\ \bibinfo
  {pages} {1--5} (\bibinfo {year} {2012})}\BibitemShut {NoStop}%
\bibitem [{\citenamefont {Rosenberg}\ \emph {et~al.}(2019)\citenamefont
  {Rosenberg}, \citenamefont {Chu}, \citenamefont {Ruff}, \citenamefont
  {Hristov},\ and\ \citenamefont {Fisher}}]{Rosenberg2019a}%
  \BibitemOpen
  \bibfield  {author} {\bibinfo {author} {\bibfnamefont {Elliott~W.}\
  \bibnamefont {Rosenberg}}, \bibinfo {author} {\bibfnamefont {Jiun~Haw}\
  \bibnamefont {Chu}}, \bibinfo {author} {\bibfnamefont {Jacob~P.C.}\
  \bibnamefont {Ruff}}, \bibinfo {author} {\bibfnamefont {Alexander~T.}\
  \bibnamefont {Hristov}}, \ and\ \bibinfo {author} {\bibfnamefont {Ian~R.}\
  \bibnamefont {Fisher}},\ }\bibfield  {title} {\enquote {\bibinfo {title}
  {{Divergence of the quadrupole-strain susceptibility of the electronic
  nematic system YbRu 2 Ge 2}},}\ }\href {\doibase 10.1073/pnas.1818910116}
  {\bibfield  {journal} {\bibinfo  {journal} {Proceedings of the National
  Academy of Sciences of the United States of America}\ }\textbf {\bibinfo
  {volume} {116}},\ \bibinfo {pages} {7232--7237} (\bibinfo {year}
  {2019})}\BibitemShut {NoStop}%
\bibitem [{\citenamefont {Freyer}\ \emph {et~al.}(2018)\citenamefont {Freyer},
  \citenamefont {Attig}, \citenamefont {Lee}, \citenamefont {Paramekanti},
  \citenamefont {Trebst},\ and\ \citenamefont {Kim}}]{Freyer2018}%
  \BibitemOpen
  \bibfield  {author} {\bibinfo {author} {\bibfnamefont {Frederic}\
  \bibnamefont {Freyer}}, \bibinfo {author} {\bibfnamefont {Jan}\ \bibnamefont
  {Attig}}, \bibinfo {author} {\bibfnamefont {Sungbin}\ \bibnamefont {Lee}},
  \bibinfo {author} {\bibfnamefont {Arun}\ \bibnamefont {Paramekanti}},
  \bibinfo {author} {\bibfnamefont {Simon}\ \bibnamefont {Trebst}}, \ and\
  \bibinfo {author} {\bibfnamefont {Yong~Baek}\ \bibnamefont {Kim}},\
  }\bibfield  {title} {\enquote {\bibinfo {title} {{Two-stage multipolar
  ordering in PrT2Al20 Kondo materials}},}\ }\href {\doibase
  10.1103/PhysRevB.97.115111} {\bibfield  {journal} {\bibinfo  {journal}
  {Physical Review B}\ }\textbf {\bibinfo {volume} {97}},\ \bibinfo {pages}
  {1--7} (\bibinfo {year} {2018})}\BibitemShut {NoStop}%
\bibitem [{\citenamefont {Lee}\ \emph {et~al.}(2018)\citenamefont {Lee},
  \citenamefont {Trebst}, \citenamefont {Kim},\ and\ \citenamefont
  {Paramekanti}}]{Lee2018b}%
  \BibitemOpen
  \bibfield  {author} {\bibinfo {author} {\bibfnamefont {Sung~Bin}\
  \bibnamefont {Lee}}, \bibinfo {author} {\bibfnamefont {Simon}\ \bibnamefont
  {Trebst}}, \bibinfo {author} {\bibfnamefont {Yong~Baek}\ \bibnamefont {Kim}},
  \ and\ \bibinfo {author} {\bibfnamefont {Arun}\ \bibnamefont {Paramekanti}},\
  }\bibfield  {title} {\enquote {\bibinfo {title} {{Landau theory of multipolar
  orders in Pr(Y)2X20 Kondo materials (Y=Ti, V, Rh, Ir; X=Al, Zn)}},}\ }\href
  {\doibase 10.1103/PhysRevB.98.134447} {\bibfield  {journal} {\bibinfo
  {journal} {Physical Review B}\ }\textbf {\bibinfo {volume} {98}},\ \bibinfo
  {pages} {134447} (\bibinfo {year} {2018})}\BibitemShut {NoStop}%
\bibitem [{\citenamefont {Patri}\ \emph {et~al.}(2019)\citenamefont {Patri},
  \citenamefont {Sakai}, \citenamefont {Lee}, \citenamefont {Paramekanti},
  \citenamefont {Nakatsuji},\ and\ \citenamefont {Kim}}]{Patri2019c}%
  \BibitemOpen
  \bibfield  {author} {\bibinfo {author} {\bibfnamefont {Adarsh~S.}\
  \bibnamefont {Patri}}, \bibinfo {author} {\bibfnamefont {Akito}\ \bibnamefont
  {Sakai}}, \bibinfo {author} {\bibfnamefont {Sung~Bin}\ \bibnamefont {Lee}},
  \bibinfo {author} {\bibfnamefont {Arun}\ \bibnamefont {Paramekanti}},
  \bibinfo {author} {\bibfnamefont {Satoru}\ \bibnamefont {Nakatsuji}}, \ and\
  \bibinfo {author} {\bibfnamefont {Yong~Baek}\ \bibnamefont {Kim}},\
  }\bibfield  {title} {\enquote {\bibinfo {title} {{Unveiling hidden multipolar
  orders with magnetostriction}},}\ }\href {\doibase
  10.1038/s41467-019-11913-3} {\bibfield  {journal} {\bibinfo  {journal}
  {Nature Communications}\ }\textbf {\bibinfo {volume} {10}} (\bibinfo {year}
  {2019}),\ 10.1038/s41467-019-11913-3}\BibitemShut {NoStop}%
\bibitem [{\citenamefont {Kuramoto}\ \emph {et~al.}(2009)\citenamefont
  {Kuramoto}, \citenamefont {Kusunose},\ and\ \citenamefont
  {Kiss}}]{Kuramoto2009}%
  \BibitemOpen
  \bibfield  {author} {\bibinfo {author} {\bibfnamefont {Yoshio}\ \bibnamefont
  {Kuramoto}}, \bibinfo {author} {\bibfnamefont {Hiroaki}\ \bibnamefont
  {Kusunose}}, \ and\ \bibinfo {author} {\bibfnamefont {Annam{\'{a}}ria}\
  \bibnamefont {Kiss}},\ }\bibfield  {title} {\enquote {\bibinfo {title}
  {{Multipole orders and fluctuations in strongly correlated electron
  systems}},}\ }\href {\doibase 10.1143/JPSJ.78.072001} {\bibfield  {journal}
  {\bibinfo  {journal} {Journal of the Physical Society of Japan}\ }\textbf
  {\bibinfo {volume} {78}},\ \bibinfo {pages} {1--33} (\bibinfo {year}
  {2009})}\BibitemShut {NoStop}%
\bibitem [{\citenamefont {Cox}(1987)}]{Cox1987}%
  \BibitemOpen
  \bibfield  {author} {\bibinfo {author} {\bibfnamefont {D.~L.}\ \bibnamefont
  {Cox}},\ }\bibfield  {title} {\enquote {\bibinfo {title} {{Quadrupolar Kondo
  effect in uranium heavy-electron materials?}}}\ }\href {\doibase
  10.1103/PhysRevLett.59.1240} {\bibfield  {journal} {\bibinfo  {journal}
  {Physical Review Letters}\ }\textbf {\bibinfo {volume} {59}},\ \bibinfo
  {pages} {1240--1243} (\bibinfo {year} {1987})}\BibitemShut {NoStop}%
\bibitem [{\citenamefont {Cox}(1988)}]{Cox1988}%
  \BibitemOpen
  \bibfield  {author} {\bibinfo {author} {\bibfnamefont {D.~L.}\ \bibnamefont
  {Cox}},\ }\bibfield  {title} {\enquote {\bibinfo {title} {{The quadrupolar
  Kondo effect: A new mechanism for heavy electrons}},}\ }\href {\doibase
  10.1016/0304-8853(88)90315-0} {\bibfield  {journal} {\bibinfo  {journal}
  {Journal of Magnetism and Magnetic Materials}\ }\textbf {\bibinfo {volume}
  {76-77}},\ \bibinfo {pages} {53--58} (\bibinfo {year} {1988})}\BibitemShut
  {NoStop}%
\bibitem [{\citenamefont {Yamane}\ \emph {et~al.}(2018)\citenamefont {Yamane},
  \citenamefont {Onimaru}, \citenamefont {Wakiya}, \citenamefont {Matsumoto},
  \citenamefont {Umeo},\ and\ \citenamefont {Takabatake}}]{Yamane2018}%
  \BibitemOpen
  \bibfield  {author} {\bibinfo {author} {\bibfnamefont {Y.}~\bibnamefont
  {Yamane}}, \bibinfo {author} {\bibfnamefont {T.}~\bibnamefont {Onimaru}},
  \bibinfo {author} {\bibfnamefont {K.}~\bibnamefont {Wakiya}}, \bibinfo
  {author} {\bibfnamefont {K.~T.}\ \bibnamefont {Matsumoto}}, \bibinfo {author}
  {\bibfnamefont {K.}~\bibnamefont {Umeo}}, \ and\ \bibinfo {author}
  {\bibfnamefont {T.}~\bibnamefont {Takabatake}},\ }\bibfield  {title}
  {\enquote {\bibinfo {title} {{Single-Site Non-Fermi-Liquid Behaviors in a
  Diluted 4f2 System Y1-xPrxIr2Zn20}},}\ }\href {\doibase
  10.1103/PhysRevLett.121.077206} {\bibfield  {journal} {\bibinfo  {journal}
  {Physical Review Letters}\ }\textbf {\bibinfo {volume} {121}},\ \bibinfo
  {pages} {77206} (\bibinfo {year} {2018})}\BibitemShut {NoStop}%
\bibitem [{\citenamefont {Yanagisawa}\ \emph {et~al.}(2020)\citenamefont
  {Yanagisawa}, \citenamefont {Hidaka}, \citenamefont {Amitsuka}, \citenamefont
  {Zherlitsyn}, \citenamefont {Wosnitza}, \citenamefont {Yamane},\ and\
  \citenamefont {Onimaru}}]{Yanagisawa2019}%
  \BibitemOpen
  \bibfield  {author} {\bibinfo {author} {\bibfnamefont {Tatsuya}\ \bibnamefont
  {Yanagisawa}}, \bibinfo {author} {\bibfnamefont {Hiroyuki}\ \bibnamefont
  {Hidaka}}, \bibinfo {author} {\bibfnamefont {Hiroshi}\ \bibnamefont
  {Amitsuka}}, \bibinfo {author} {\bibfnamefont {Sergei}\ \bibnamefont
  {Zherlitsyn}}, \bibinfo {author} {\bibfnamefont {Joachim}\ \bibnamefont
  {Wosnitza}}, \bibinfo {author} {\bibfnamefont {Yu}~\bibnamefont {Yamane}}, \
  and\ \bibinfo {author} {\bibfnamefont {Takahiro}\ \bibnamefont {Onimaru}},\
  }\bibfield  {title} {\enquote {\bibinfo {title} {{Logarithmic Elastic
  Response in the Dilute Non-Kramers System Y 1? x Pr x Ir 2 Zn 20}},}\ \
  }(\bibinfo  {publisher} {Journal of the Physical Society of Japan},\ \bibinfo
  {year} {2020})\ pp.\ \bibinfo {pages} {1--9}\BibitemShut {NoStop}%
\bibitem [{\citenamefont {Araki}\ \emph {et~al.}(2014)\citenamefont {Araki},
  \citenamefont {Shimura}, \citenamefont {Kase}, \citenamefont {Sakakibara},
  \citenamefont {Sakai},\ and\ \citenamefont {Nakatsuji}}]{Araki2014}%
  \BibitemOpen
  \bibfield  {author} {\bibinfo {author} {\bibfnamefont {K.}~\bibnamefont
  {Araki}}, \bibinfo {author} {\bibfnamefont {Y.}~\bibnamefont {Shimura}},
  \bibinfo {author} {\bibfnamefont {N.}~\bibnamefont {Kase}}, \bibinfo {author}
  {\bibfnamefont {T.}~\bibnamefont {Sakakibara}}, \bibinfo {author}
  {\bibfnamefont {A.}~\bibnamefont {Sakai}}, \ and\ \bibinfo {author}
  {\bibfnamefont {S.}~\bibnamefont {Nakatsuji}},\ }\bibfield  {title} {\enquote
  {\bibinfo {title} {{Magnetization and Specific Heat of the Cage Compound PrV
  2 Al 20}},}\ \ }(\bibinfo  {publisher} {Journal of the Physical Society of
  Japan},\ \bibinfo {year} {2014})\ pp.\ \bibinfo {pages} {3--7}\BibitemShut
  {NoStop}%
\bibitem [{\citenamefont {Patri}\ \emph {et~al.}(2020)\citenamefont {Patri},
  \citenamefont {Khait},\ and\ \citenamefont {Kim}}]{Patri2020b}%
  \BibitemOpen
  \bibfield  {author} {\bibinfo {author} {\bibfnamefont {Adarsh~S.}\
  \bibnamefont {Patri}}, \bibinfo {author} {\bibfnamefont {Ilia}\ \bibnamefont
  {Khait}}, \ and\ \bibinfo {author} {\bibfnamefont {Yong~Baek}\ \bibnamefont
  {Kim}},\ }\bibfield  {title} {\enquote {\bibinfo {title} {{Emergent
  non-Fermi-liquid phenomena in multipolar quantum impurity systems}},}\ }\href
  {\doibase 10.1103/PhysRevResearch.2.013257} {\bibfield  {journal} {\bibinfo
  {journal} {Physical Review Research}\ }\textbf {\bibinfo {volume} {2}},\
  \bibinfo {pages} {013257} (\bibinfo {year} {2020})}\BibitemShut {NoStop}%
\bibitem [{\citenamefont {Patri}\ and\ \citenamefont {Kim}(2020)}]{Patri2020c}%
  \BibitemOpen
  \bibfield  {author} {\bibinfo {author} {\bibfnamefont {Adarsh~S.}\
  \bibnamefont {Patri}}\ and\ \bibinfo {author} {\bibfnamefont {Yong~Baek}\
  \bibnamefont {Kim}},\ }\bibfield  {title} {\enquote {\bibinfo {title}
  {{Critical Theory of Non-Fermi Liquid Fixed Point in Multipolar Kondo
  Problem}},}\ }\href {\doibase 10.1103/PhysRevX.10.041021} {\bibfield
  {journal} {\bibinfo  {journal} {Physical Review X}\ }\textbf {\bibinfo
  {volume} {10}},\ \bibinfo {pages} {41021} (\bibinfo {year}
  {2020})}\BibitemShut {NoStop}%
\bibitem [{\citenamefont {Onimaru}\ and\ \citenamefont
  {Kusunose}(2016)}]{Onimaru2016}%
  \BibitemOpen
  \bibfield  {author} {\bibinfo {author} {\bibfnamefont {Takahiro}\
  \bibnamefont {Onimaru}}\ and\ \bibinfo {author} {\bibfnamefont {Hiroaki}\
  \bibnamefont {Kusunose}},\ }\bibfield  {title} {\enquote {\bibinfo {title}
  {{Exotic quadrupolar phenomena in non-Kramers doublet systems - The cases of
  PrT2Zn20 (T = Ir, Rh) and PrT2Al20 (T = V, Ti)}},}\ }\href {\doibase
  10.7566/JPSJ.85.082002} {\bibfield  {journal} {\bibinfo  {journal} {Journal
  of the Physical Society of Japan}\ }\textbf {\bibinfo {volume} {85}},\
  \bibinfo {pages} {1--22} (\bibinfo {year} {2016})}\BibitemShut {NoStop}%
\bibitem [{\citenamefont {Nagashima}\ \emph {et~al.}(2014)\citenamefont
  {Nagashima}, \citenamefont {Nishiwaki}, \citenamefont {Otani}, \citenamefont
  {Sakoda}, \citenamefont {Matsuoka}, \citenamefont {Harima},\ and\
  \citenamefont {Sugawara}}]{Nagashima2014}%
  \BibitemOpen
  \bibfield  {author} {\bibinfo {author} {\bibfnamefont {Souta}\ \bibnamefont
  {Nagashima}}, \bibinfo {author} {\bibfnamefont {Taihei}\ \bibnamefont
  {Nishiwaki}}, \bibinfo {author} {\bibfnamefont {Akira}\ \bibnamefont
  {Otani}}, \bibinfo {author} {\bibfnamefont {Masahito}\ \bibnamefont
  {Sakoda}}, \bibinfo {author} {\bibfnamefont {Eiichi}\ \bibnamefont
  {Matsuoka}}, \bibinfo {author} {\bibfnamefont {Hisatomo}\ \bibnamefont
  {Harima}}, \ and\ \bibinfo {author} {\bibfnamefont {Hitoshi}\ \bibnamefont
  {Sugawara}},\ }\bibfield  {title} {\enquote {\bibinfo {title} {{De Haas?Van
  Alphen Effect in RTi 2 Al 20 (R = La, Pr, and Sm)}},}\ \ }(\bibinfo
  {publisher} {Journal of the Physical Society of Japan},\ \bibinfo {year}
  {2014})\ pp.\ \bibinfo {pages} {2--7}\BibitemShut {NoStop}%
\bibitem [{\citenamefont {Wilson}(1975)}]{Wilson1975a}%
  \BibitemOpen
  \bibfield  {author} {\bibinfo {author} {\bibfnamefont {Kenneth~G.}\
  \bibnamefont {Wilson}},\ }\bibfield  {title} {\enquote {\bibinfo {title}
  {{The renormalization group: Critical phenomena and the Kondo problem}},}\
  }\href {\doibase 10.1103/RevModPhys.47.773} {\bibfield  {journal} {\bibinfo
  {journal} {Reviews of Modern Physics}\ }\textbf {\bibinfo {volume} {47}},\
  \bibinfo {pages} {773--840} (\bibinfo {year} {1975})}\BibitemShut {NoStop}%
\bibitem [{\citenamefont {Affleck}\ and\ \citenamefont
  {Ludwig}(1993)}]{Affleck1993a}%
  \BibitemOpen
  \bibfield  {author} {\bibinfo {author} {\bibfnamefont {Ian}\ \bibnamefont
  {Affleck}}\ and\ \bibinfo {author} {\bibfnamefont {Andreas W~W}\ \bibnamefont
  {Ludwig}},\ }\bibfield  {title} {\enquote {\bibinfo {title} {{Exact
  conformal-field-theory results on the multichannel Kondo effect:
  Single-fermion Green's function, self-energy, and resistivity}},}\ }\href
  {\doibase 10.1103/PhysRevB.48.7297} {\bibfield  {journal} {\bibinfo
  {journal} {Physical Review B}\ }\textbf {\bibinfo {volume} {48}},\ \bibinfo
  {pages} {7297--7321} (\bibinfo {year} {1993})}\BibitemShut {NoStop}%
\bibitem [{\citenamefont {Affleck}\ and\ \citenamefont
  {Ludwig}(1991)}]{Affleck1991}%
  \BibitemOpen
  \bibfield  {author} {\bibinfo {author} {\bibfnamefont {Ian}\ \bibnamefont
  {Affleck}}\ and\ \bibinfo {author} {\bibfnamefont {Andreas~W.W.}\
  \bibnamefont {Ludwig}},\ }\bibfield  {title} {\enquote {\bibinfo {title}
  {{Critical theory of overscreened Kondo fixed points}},}\ }\href {\doibase
  10.1016/0550-3213(91)90419-X} {\bibfield  {journal} {\bibinfo  {journal}
  {Nuclear Physics, Section B}\ }\textbf {\bibinfo {volume} {360}},\ \bibinfo
  {pages} {641--696} (\bibinfo {year} {1991})}\BibitemShut {NoStop}%
\bibitem [{\citenamefont {Sakai}\ and\ \citenamefont
  {Nakatsuji}(2011)}]{Sakai2011a}%
  \BibitemOpen
  \bibfield  {author} {\bibinfo {author} {\bibfnamefont {Akito}\ \bibnamefont
  {Sakai}}\ and\ \bibinfo {author} {\bibfnamefont {Satoru}\ \bibnamefont
  {Nakatsuji}},\ }\bibfield  {title} {\enquote {\bibinfo {title} {{Kondo
  effects and multipolar order in the cubic PrTr2Al 20 (Tr = Ti, V)}},}\ }\href
  {\doibase 10.1143/JPSJ.80.063701} {\bibfield  {journal} {\bibinfo  {journal}
  {Journal of the Physical Society of Japan}\ }\textbf {\bibinfo {volume}
  {80}},\ \bibinfo {pages} {1--4} (\bibinfo {year} {2011})}\BibitemShut
  {NoStop}%
\bibitem [{\citenamefont {Wugalter}\ \emph {et~al.}(2020)\citenamefont
  {Wugalter}, \citenamefont {Komijani},\ and\ \citenamefont
  {Coleman}}]{Wugalter2020}%
  \BibitemOpen
  \bibfield  {author} {\bibinfo {author} {\bibfnamefont {Ari}\ \bibnamefont
  {Wugalter}}, \bibinfo {author} {\bibfnamefont {Yashar}\ \bibnamefont
  {Komijani}}, \ and\ \bibinfo {author} {\bibfnamefont {Piers}\ \bibnamefont
  {Coleman}},\ }\bibfield  {title} {\enquote {\bibinfo {title} {{Large-N
  approach to the two-channel Kondo lattice}},}\ }\href {\doibase
  10.1103/PhysRevB.101.075133} {\bibfield  {journal} {\bibinfo  {journal}
  {Physical Review B}\ }\textbf {\bibinfo {volume} {101}},\ \bibinfo {pages}
  {1--20} (\bibinfo {year} {2020})}\BibitemShut {NoStop}%
\bibitem [{\citenamefont {Inui}\ and\ \citenamefont {Motome}(2020)}]{Inui2020}%
  \BibitemOpen
  \bibfield  {author} {\bibinfo {author} {\bibfnamefont {Koji}\ \bibnamefont
  {Inui}}\ and\ \bibinfo {author} {\bibfnamefont {Yukitoshi}\ \bibnamefont
  {Motome}},\ }\bibfield  {title} {\enquote {\bibinfo {title}
  {{Channel-selective non-Fermi liquid behavior in the two-channel Kondo
  lattice model under a magnetic field}},}\ }\href {\doibase
  10.1103/PhysRevB.102.155126} {\bibfield  {journal} {\bibinfo  {journal}
  {Physical Review B}\ }\textbf {\bibinfo {volume} {102}} (\bibinfo {year}
  {2020}),\ 10.1103/PhysRevB.102.155126}\BibitemShut {NoStop}%
\bibitem [{\citenamefont {Lai}\ \emph {et~al.}(2018)\citenamefont {Lai},
  \citenamefont {Nica}, \citenamefont {Hu}, \citenamefont {Gong}, \citenamefont
  {Paschen},\ and\ \citenamefont {Si}}]{Lai2018}%
  \BibitemOpen
  \bibfield  {author} {\bibinfo {author} {\bibfnamefont {Hsin-Hua}\
  \bibnamefont {Lai}}, \bibinfo {author} {\bibfnamefont {Emilian~M.}\
  \bibnamefont {Nica}}, \bibinfo {author} {\bibfnamefont {Wen-Jun}\
  \bibnamefont {Hu}}, \bibinfo {author} {\bibfnamefont {Shou-Shu}\ \bibnamefont
  {Gong}}, \bibinfo {author} {\bibfnamefont {Silke}\ \bibnamefont {Paschen}}, \
  and\ \bibinfo {author} {\bibfnamefont {Qimiao}\ \bibnamefont {Si}},\
  }\bibfield  {title} {\enquote {\bibinfo {title} {{Kondo Destruction and
  Multipolar Order-- Implications for Heavy Fermion Quantum Criticality}},}\
  }\href {http://arxiv.org/abs/1807.09258} {\  (\bibinfo {year} {2018})},\
  \Eprint {http://arxiv.org/abs/1807.09258} {arXiv:1807.09258} \BibitemShut
  {NoStop}%
\bibitem [{\citenamefont {Hu}\ \emph {et~al.}(2020)\citenamefont {Hu},
  \citenamefont {Cai},\ and\ \citenamefont {Si}}]{Hu2020}%
  \BibitemOpen
  \bibfield  {author} {\bibinfo {author} {\bibfnamefont {Haoyu}\ \bibnamefont
  {Hu}}, \bibinfo {author} {\bibfnamefont {Ang}\ \bibnamefont {Cai}}, \ and\
  \bibinfo {author} {\bibfnamefont {Qimiao}\ \bibnamefont {Si}},\ }\bibfield
  {title} {\enquote {\bibinfo {title} {{Quantum Criticality and Dynamical Kondo
  Effect in an SU(2) Anderson Lattice Model}},}\ }\href@noop {} {\bibfield
  {journal} {\bibinfo  {journal} {arXiv}\ }\textbf {\bibinfo {volume} {2}}
  (\bibinfo {year} {2020})},\ \Eprint {http://arxiv.org/abs/2004.04679}
  {arXiv:2004.04679} \BibitemShut {NoStop}%
\bibitem [{\citenamefont {Nozieres}\ and\ \citenamefont
  {Blandin}(1980)}]{Nozieres1980}%
  \BibitemOpen
  \bibfield  {author} {\bibinfo {author} {\bibfnamefont {Ph}~\bibnamefont
  {Nozieres}}\ and\ \bibinfo {author} {\bibfnamefont {A.}~\bibnamefont
  {Blandin}},\ }\bibfield  {title} {\enquote {\bibinfo {title} {{Kondo Effect
  in Real Metals.}}}\ }\href {\doibase 10.1051/jphys:01980004103019300}
  {\bibfield  {journal} {\bibinfo  {journal} {Journal de physique Paris}\
  }\textbf {\bibinfo {volume} {41}},\ \bibinfo {pages} {193--211} (\bibinfo
  {year} {1980})}\BibitemShut {NoStop}%
\end{thebibliography}
\end{document}